\documentclass{article}

\usepackage{fullpage}
\usepackage{hyperref}
\usepackage{amsmath}
\usepackage{amssymb}
\usepackage{amsthm}
\usepackage{color}
\usepackage{authblk}
\usepackage{comment}

\newtheorem{proposition}{Proposition}

\newtheorem{theorem}{Theorem}
\newtheorem{definition}{Definition}
\newtheorem{corollary}{Corollary}
\newtheorem{remark}{Remark}

\numberwithin{proposition}{section}
\numberwithin{lemma}{section}
\numberwithin{theorem}{section}
\numberwithin{definition}{section}
\numberwithin{corollary}{section}
\numberwithin{remark}{section}

\def\Real{\textnormal{Re}}

\def\LWE{\textnormal{LWE}}
\def\NFLWE{\textnormal{NF-LWE}}

\def\GenTrap{\textnormal{GenTrap}}
\def\Invert{\textnormal{Invert}}
\def\Encrypt{\textnormal{Encrypt}}
\def\FakeEncrypt{\textnormal{FakeEncrypt}}
\def\Score{\textnormal{Score}}
\def\DecodeError{\textnormal{DecodeError}}

\def\negl{\textnormal{negl}}

\def\Tr{\textnormal{Tr}}

\def\Supp{\textnormal{Supp }}

\def\Image{\textnormal{Image}}

\def\FirstResponse{\textnormal{FirstResponse}}
\def\SecondResponse{\textnormal{SecondResponse}}

\def\BestScore{\textnormal{BestScore}}

\date{}

\begin{document}

\title{Hidden-State Proofs of Quantumness}

\author{Carl A.~Miller}

\affil{\small Joint Center for Quantum Information and Computer Science (QuICS) \\  3100 Atlantic Building, College Park, MD 20742  \\  \vskip0.1in
National Institute of Standards and Technology (NIST) \\  100 Bureau Dr., Gaithersburg, MD 20899  
}

\maketitle

\begin{abstract}
An experimental cryptographic proof of quantumness will be a vital milestone in the progress of quantum information science.  Error tolerance is a persistent challenge for implementing such tests: we need a test that not only can be passed by an efficient quantum prover, but one that can be passed by a prover that exhibits a certain amount of computational error.  (Brakerski et al.~2018) introduced an innovative two-round proof of quantumness based on the  Learning With Errors (LWE) assumption.  However, one of the steps in their protocol (the pre-image test) has low tolerance for error.  In this work we present a proof of quantumness which maintains the same circuit structure as (Brakerski et al.~2018)  while improving the robustness for noise.  Our protocol is based on cryptographically hiding an extended Greenberger-Horne-Zeilinger (GHZ) state within a sequence of classical bits.  Asymptotically, our protocol allows the total probability of error within the circuit to be as high as $1 - O ( \lambda^{-C} )$, where $\lambda$ is the security parameter and $C$ is a constant that can be made arbitrarily large.  As part of the proof of this result, we also prove an uncertainty principle over finite abelian groups which may be of independent interest.
\end{abstract}

\section{Introduction}

As advances in quantum computing continue to accelerate, a central question is: how will we know when we have a quantum advantage in computing?  Will we be able to prove that  such an advantage has been achieved?  The highest-profile target for quantum computing continues to be implementing Shor's algorithm \cite{shor1999polynomial}, but that target remains a distant goal, in large part because of the need for quantum error-correction.  Is it possible to prove the quantum behavior of a computer in the absence of full quantum error correction?

Google's breakthrough demonstration on a $53$-qubit  device \cite{arute2019quantum} was based on random circuit sampling, and it had a high tolerance for error. However, Google's claim that \cite{arute2019quantum} was a demonstration of quantum advantage was based on assumptions about the hardness of classically simulating quantum circuits, and those assumptions were (fairly drastically) disproved after the fact \cite{cho2022ordinary}.  Attempts to prove quantumness based on instantaneous quantum polynomial time (IQP) computation \cite{shepherd2009temporally} have similarly encountered challenges to the underlying assumptions \cite{kahanamoku2023forging}.
While advances on these topics continue, it is useful to consider whether another direction can provide more stable claims of a computational quantum advantage.

In a breakthrough result in 2018, Brakerski et al.~\cite{brakerski2021cryptographic} established a theoretical interactive proof of quantumness based on the LWE (Learning With Errors) assumption.  A benefit of using LWE as a starting point is that it has a long history as a basis for classical cryptographic protocols \cite{regev2009lattices}, and thus there are well-established and stable metrics for the hardness of LWE problems.  The protocol of \cite{brakerski2021cryptographic} begins by having a classical verifier randomly construct a pair of linear injective functions
\begin{eqnarray}
    f_0, f_1 \colon \mathbb{Z}_q^n \to \mathbb{Z}_q^m
\end{eqnarray}
which have approximately the same image,
in the sense that for any $\mathbf{s}_0 \in \mathbb{Z}_q^n$, there exists $\mathbf{s}_1 \in \mathbb{Z}_q^n$ such that the difference vector $f_0 ( \mathbf{s}_0 ) - f_1 ( \mathbf{s}_1 ) $ has entries close to $0$. However, the LWE assumption implies that it is impossible for any outside party to reliably compute such a pair $(\mathbf{s}_0, \mathbf{s}_1)$.

In the first round of the protocol, the verifier has the quantum prover use $f_0, f_1$ to prepare a \textit{claw-state}
\begin{eqnarray}
    \label{initclaw}
    \phi & = & \frac{1}{\sqrt{2}} \left( \left| \mathbf{s}_0 , 0 \right> + \left| \mathbf{s}_1, 1 \right> \right).
\end{eqnarray}
Then, in the second round of the protocol, according to a random coin flip, the verifier either asks the prover to measure $\phi$ in the computational basis and report the result (the ``pre-image test''), or to measure in the Hadamard basis and report the result (the ``equation test'').  The verifier (who knows a classical description of the state $\phi$) checks whether the result was valid; if so, the prover passes.  Otherwise, the prover fails.

A quantum prover can (in theory) pass this test with probability approaching $1$.  However, a classical prover cannot do better than $\frac{3}{4} + \negl(\lambda)$, where $\lambda$ denotes the security parameter.  This fact is proved by a rewinding argument: if a classical prover can pass the pre-image and equation tests with high probability, then the same prover could pass both tests simultaneously, and that would mean they could compute information information about $\mathbf{s}_0, \mathbf{s}_1$ beyond what the LWE assumption allows.  Therefore, as long as the LWE problem cannot be solved in polynomial time, the protocol is sound against classical provers.  

However, error-tolerance continues to be an issue.  The pre-image test requires the prover to report correctly either the entire vector $\mathbf{s}_0$ or the entire vector $\mathbf{s}_1$.  Naively, if the probability of an error occurring before or during the pre-image test is significantly more than $1/2$, then the prover's expected score will be below $(1/2)(1/2) + (1/2)(1) = (3/4)$, and the prover will fail too often to achieve a quantum advantage.

Our goal in this paper is to improve the error tolerance of  protocols in the form of \cite{brakerski2021cryptographic}.  We begin by considering a parallel research topic (nonlocal games).

\subsection{Interactive proofs of quantumness versus nonlocal games}

A nonlocal game is a different type of proof of quantumness with a much longer history, going back at least to the work of John Bell in the 1960s. In this setting, a classical verifier interacts with $2$ or more provers.  The verifier assigns a single score to the prover at the end of the interaction. 
If the average score of the provers (across multiple rounds) lies significantly outside the range achievable by classical provers, the verifiers conclude that the provers must be quantum.

A simple way to measure the error-tolerance of a nonlocal game is via the \textbf{bias ratio}.  If $H$ is a nonlocal game, the bias ratio is given by the expression
\begin{eqnarray}
    r ( H ) & = & \frac{ \sup_{Q} \left| \omega ( H, Q ) \right| }{ \sup_{C} \left| \omega ( H, C ) \right| },
    \label{theratio}
\end{eqnarray}
where $C$ varies over all classical strategies for $H$ and $Q$ varies over all quantum strategies for $H$, and $\omega ( H, S )$ denotes the expected score of a given strategy $S$.  (See Section III.B.2 of \cite{brunner2014bell} for a discussion about this metric.)
The numerator in equation~(\ref{theratio}) is the \textbf{quantum bias}, and the denominator is the \textbf{classical bias}.  If one accepts the heuristic that when an error occurs, the provers will perform no worse than the worst classical score (i.e., they do not accidentally achieve a Bell violation in the opposite direction), then we find that the provers can prove quantum behavior as long as their total error probability is less than  $(r(H)-1)/(r(H)+1) = 1 - O ( 1 / r ( H ))$.  Thus, a large bias ratio is desirable.

We consider previous work under this metric.
For the original test from \cite{brakerski2021cryptographic}, let us suppose that a score of $+1$ is awarded when the prover passes, and a score of $-1$ is awarded when the prover fails.  
A classical prover can easily achieve an expected score as high as $1/2$ or as low as $-1/2$.
Since a quantum prover's score is restricted to $[-1,1]$, this implies a bias ratio of at most $2$ for the test in \cite{brakerski2021cryptographic}.   
Subsequent work in \cite{kahanamoku2022classically} and \cite{alnawakhtha2024lattice} modified and optimized the test in \cite{brakerski2021cryptographic}. However, the tests from both \cite{kahanamoku2022classically} and \cite{alnawakhtha2024lattice} have bias ratios significantly less than $2$.  For the proof of quantumness in \cite{alnawakhtha2024lattice}, the range of classical winning probabilities is at least $[0.25, 0.75]$ and the proved range of quantum winning probabilities is approximately $[ 0.1464, 0.8536 ]$.  For \cite{kahanamoku2022classically}, if one uses the expression on the left side of inequality (1) in Theorem 2 as the scoring function, then the classical range of expected scores at least $[-3,0]$ and the quantum range of expected scores is at most $[-4, 1]$.\footnote{We note that the authors in \cite{kahanamoku2022classically} also claim to give techniques that improve the error-tolerance of their proof of quantumness to a much higher degree.  However, this claim is based on a restricted error model which we do not wish to assume in this paper.  Their error model assumes that a noisy claw-state has a particular form, given in equation (1) in the Supplementary Information of \cite{kahanamoku2022classically}.  This assumption does not allow for the possibility that non-real complex phases could occur in a noisy claw-state.}

In \cite{kalai2023quantum}, a general compiler was given that can convert any multi-player nonlocal game into an interactive proof of quantumness based on LWE.  This result suggests taking a known nonlocal game with very large bias ratio (such as one of the games from \cite{buhrman2011near}) and compiling it with \cite{kalai2023quantum} to obtain an interactive proof of quantumness with very large bias ratio.  This approach is viable, but there is an important catch: \cite{kalai2023quantum} requires using the full machinery of quantum homomorphic encryption, which means substantially increasing the quantum circuits involved in the interactive test and thereby increasing the probability of an error.  The motivating question behind the current work is the following: Can we, without changing the basic quantum circuit structure of \cite{brakerski2021cryptographic}, create a modified protocol in which the bias ratio tends to infinity?

\subsection{Main Result}

One of the oldest known families of nonlocal games with bias ratio tending to infinity is the family of extended GHZ games.  The $GHZ_k$ game is played by $k$-players Alice$_1$, Alice$_2$, $\ldots$, Alice$_k$, with $k \geq 3$.  A referee randomly chooses a bit string $\mathbf{x} = x_1 x_2 \ldots x_k$ of even parity and sends $x_j$ to Alice$_j$ for each $j$.  Each Alice$_j$ then outputs a bit $a_j$.  The players score $+1$ if the following equation holds:
\begin{eqnarray}
    x_1 + x_2 + \ldots + x_k + 2 a_1 + 2a_2 + \ldots + 2a_k & \equiv & 0 \hskip0.1in (\textnormal{mod } 4)
\end{eqnarray}
and they score $-1$ if it does not hold.
This game can be won perfectly if the players share a $k$-qubit GHZ state
$\frac{1}{\sqrt{2}} \left( \left| 0^k \right> + \left| 1^k \right> \right)$ and Alice$_j$ applies a Pauli $X$-measurement on her qubit if $x_j = 0$ or a Pauli $Y$-measurement on her qubit if $x_j = 1$.  However, a result by Mermin \cite{mermin1990extreme} implies that classical players will always have an average score between $-2^{-(k/2)+1}$ and $2^{-(k/2)+1}$,  implying that bias ratio for the extended GHZ games tends to infinity at an exponential rate.

Our main result, building on \cite{brakerski2021cryptographic,kahanamoku2022classically,kalai2023quantum,alnawakhtha2024lattice} is an efficient way of compiling the extended GHZ game into an interactive proof of quantumness.
The proof of quantumness is called Game $\mathbf{R}$ in this paper, and is shown in Figure~\ref{fig:poq}.  
The first round of the protocol is modified: the functions $f_0,f_1$ are chosen somewhat differently by the verifier, and some of the qubits of the claw-state (\ref{initclaw}) are measured immediately so that before the second round, the prover holds a state of the form
\begin{eqnarray}
    \label{phiprime}
    \phi' & = & \frac{1}{\sqrt{2}} \left( \left| \mathbf{c} , 0 \right> \pm \left| \mathbf{c}', 1 \right> \right).
\end{eqnarray}
where $\mathbf{c}, \mathbf{c}'$ are bit strings of length $d \leq n$.  Crucially, the XOR string $\mathbf{c} \oplus \mathbf{c}'$ is cryptographically hidden from the prover.

In the second round, the verifier has the prover measure the first $d$ qubits of $\phi'$ randomly in either the $X$- or $Y$-bases.  The prover measures the $(d+1)$th qubit in the $(X+Y)/\sqrt{2}$-basis.  All results are reported to the verifier, who checks a certain parity condition, and awards a score of $+1$ if the parity condition is satisfied and $-1$ if it is not. 

Game $\mathbf{R}'$ (Figure~\ref{fig:poqprime}) is the same as Game $\mathbf{R}$ except that the measurements of the qubits of the state $\phi'$ are done sequentially by the prover, making Game $\mathbf{R}$ a $(d+2)$-round protocol instead of a $2$-round protocol.  We prove the following (see Propositions \ref{prop:quantumgamer}, \ref{prop:quantumgamerprime} and Theorems~\ref{thm:classicalgamer}, \ref{thm:classicalgamerprime}).  
\begin{theorem}
\label{thm:summary}
    Suppose that the LWE problem is hard. 
 Then, Games $\mathbf{R}$ and $\mathbf{R}'$  satisfy the following:
\begin{enumerate}
    \item The quantum biases for Games $\mathbf{R}$ and $\mathbf{R}'$ are at least $\sqrt{2}/2 - o ( 1 )$.  

    \item The classical bias for Game $\mathbf{R}$ is at most $\exp ( - \Omega ( d )) + \negl ( \lambda )$.

    \item The classical bias for Game $\mathbf{R'}$ is at most $2 \cdot (0.75)^{d/4} + \negl ( \lambda )$.
\end{enumerate}
\end{theorem}
These results imply that the bias ratio for Game $\mathbf{R}$ is at least exponential in $d$.  The same is true for Game $\mathbf{R}'$, which has the advantage of an explicit exponential base (namely, $(0.75)^{-1/4} \approx 1.074$).  Both games maintain the basic structure of the test from \cite{brakerski2021cryptographic}: the prover prepares a single claw-state, and then measures it using single-qubit measurements.

In order for the proof method to work, the parameter $d$ must be $O ( \log \lambda)$.  This allows us, for example, to set $d = \lfloor C \log \lambda \rfloor$ for any chosen  $C > 0$, yielding a bias ratio for Game $\mathbf{R}$ that grows at a rate of $\lambda^{\Omega ( C)}$.  Thus, any polynomial rate for the bias ratio is asymptotically achievable.

\subsection{Proof Techniques}

The intuition behind Game $\mathbf{R}$ is that the verifier has hidden a GHZ state in the state (\ref{phiprime}).  The $j$th qubit is part of the GHZ state if $c_j \neq c'_j$, whereas if $c_j = c'_j$ the the $j$th qubit simply in a computational basis state (a decoy state).  The prover is compelled to play a (modified version of) the extended GHZ game on the entangled qubits without knowing where they are.  This cryptographic hiding mechanism is designed to foil coordinated classical cheating strategies by the prover.
However, a method to prove that this cryptographic hiding actually works is not so obvious, and we take a somewhat indirect approach.

We design a two-player called Game $\mathbf{J}_d$ (Figure~\ref{fig:gamej}), which is essentially the GHZ game played with decoy states as described above.  We characterize classical strategies and show that the expected score of a classical strategy can be conveniently expressed in terms of the discrete Fourier transform over $\mathbb{Z}_4^n$ (see Subsection~\ref{subsec:classval}).  

Along the way, we prove a strong uncertainty principle for the Fourier transform for finite abelian groups.  A well-known result by Donoho and Stark \cite{donoho1989uncertainty} asserts that if $h \colon G \to \mathbb{C}$ is a nonzero function on a finite abelian group $G$, and $\hat{h}$ is its Fourier transform, then
\begin{eqnarray}
    \label{donohoprelim}
    \left| \Supp h \right| \cdot \left| \Supp \hat{h} \right| & \geq & \left| G \right|
\end{eqnarray}
(Informally, this means that is is not possible for both $h$ and $\hat{h}$ to have small support, in analogy to the Heisenberg uncertainty principle.)  In Section~\ref{sec:uncertainty}, we take this result a step further by characterizing  the cases in which inequality (\ref{donohoprelim}) is close to being an equality.  We define two quantities, the \textbf{uniformity coefficient} of $h$, denoted $\nu ( h )$, and the \textbf{linearity coefficient} of $h$, denoted $\eta (h )$.  Each of these quantities is bounded between $0$ and $1$.  Roughly speaking, $\nu (h )$ measures how close $|h|$ is to being constant on its support set $\Supp h$, and $\eta ( h )$ measures how close $\Supp h$ is to being a coset of a subgroup of $G$.  We prove the following:
\begin{theorem}
\label{thm:prelimuncertainty}
Let $G$ be a finite abelian group and let $h \colon G \to \mathbb{C}$ be a nonzero function.  Then,
\begin{eqnarray}
    \frac{\left| \Supp h \right| \left| \Supp \hat{h} \right| \nu ( h ) \eta ( h ) }{\left| G \right|} \geq 1.
\end{eqnarray}
\end{theorem}
In informal terms, this theorem asserts that if inequality (\ref{donohoprelim}) is close to an equality, then $|h|$ is close to being a constant function on an affine linear subset of $G$.

As noted above, the classical bias of Game $\mathbf{J}_d$ can be expressed in terms of the discrete Fourier transform, and it turns out that the functions in this expression have very nonlinear support.  This is sufficient to imply, using the linearity coefficient $\eta ( \cdot )$, that the classical bias of the Game $\mathbf{J}_d$ vanishes exponentially in $d$ (see Section~\ref{sec:clawgame}).  Proving the nonlinearity property needed here ultimately relies not on the properties of the $d$-player game $GHZ_d$, but of the parallel repeated $4$-player game $GHZ_4^d$. We make use of a recent result \cite{braverman2023parallel} that implies that the classical winning probability of $GHZ_4^d$ vanishes exponentially in $d$.

Modifying reasoning from \cite{kalai2023quantum}, we show in Section~\ref{sec:pfs} that 
the upper bound on the classical bias of Game $\mathbf{J}_d$ implies a similar upper bound on the classical bias of Game $\mathbf{R}$.  The argument requires rewinding a classical algorithm exponentially many times in the parameter $d$.  This rewinding must be done in polynomial time, and that is why there is a need to make $d$ bounded by $O ( \log \lambda)$.

\subsection{Related Work}

The aforementioned papers \cite{brakerski2021cryptographic, kahanamoku2022classically, kalai2023quantum, alnawakhtha2024lattice} are the main predecessors to this work.
A number of other interesting works have addressed the efficiency of cryptographic proofs of quantumness from different angles.  Our focus in this paper is on the bias ratio and the quantum circuit size of the test, but one can study other metrics instead.  \cite{liu2022depth} and \cite{hirahara2021test} both studied optimizing the quantum circuit \textit{depth} of protocols in the style of \cite{brakerski2021cryptographic}.
Also, researchers have explored efficient proofs of quantumness based on cryptographic assumptions other than LWE.  Two important examples are RLWE (Ring Learning With Errors) \cite{brakerski2023simple}, and the classical hardness of factoring \cite{kahanamoku2022classically}, both of which are standard cryptographic assumptions. Integrating the results in this paper with RLWE or factoring is a good topic for further research. 
 The paper \cite{brakerski2020simpler} established a lattice-based proof of quantumness in a single round by additionally assuming the random oracle model (ROM) assumption for a hash function.  (Integrating hash functions into the protocols given here also appears to be a good direction, although one would have to accept additional circuit complexity and the fact that the random oracle model is not a literally true assumption.)  Other, less standard cryptographic assumptions have also been explored \cite{alamati2022candidate, arabadjieva2024single, morimae2023proofs}.

A prototype experimental proof of quantumness, following \cite{brakerski2021cryptographic, kahanamoku2022classically}, was performed and reported  in \cite{zhu2023interactive}.

\subsection{Future Directions}

A natural next step is to calculate how effectively Game $\mathbf{R}$ performs as a proof of quantumness at specific values for the parameters ($n$, $q$, etc.). In Section~\ref{sec:numerical}, we give a preliminary example calculation showing how this can be done.   The proof of Theorem~\ref{thm:summary} shows that if a classical adversary can achieve a certain score at Game $\mathbf{R}$ or $\mathbf{R}'$, then the same algorithms used by the adversary can also be used to attack the LWE problem.  In Section~\ref{sec:numerical} we show that a score of at least $0.1617$ at Game $\mathbf{R}'$ implies an attack on the LWE problem.  Since a quantum adversary can win Game $\mathbf{R}'$ with probability close to $\sqrt{2} / 2 \approx 0.7071$, this suggests a quantum-to-classical bias ratio of at least roughly $0.7071 / 0.1617 \approx 4.372$.  (We note that this figure may not be directly comparable to the aforementioned bias ratios from \cite{kahanamoku2022classically, alnawakhtha2024lattice} --- our implied attack on LWE is more computationally expensive than those in \cite{kahanamoku2022classically, alnawakhtha2024lattice}, and therefore may require a higher range of LWE parameters.)

Further optimization to the tools in this paper could yield better figures.  In particular, proving variants and optimizations of the uncertainty principle (Theorem~\ref{thm:prelimuncertainty}) will translate into improvements on the bias ratio for Games $\mathbf{R}$ and $\mathbf{R}'$.  

\subsection{Acknowledgements}

Thanks to Yusuf Alnawakhtha, Greg Kahanamoku-Meyer, Khalil Guy, Yi-Kai Liu, Serge Massar, Chris Monroe,  Ryan O'Donnell, and Rene Peralta for their help with this paper.  This work was supported by the National Insitute of Standards and Technology.  The opinions expressed are solely those of the author.

\section{Preliminaries}

\label{sec:prelim}

For any positive integer $q$, let $\mathbb{Z}_q$ denote the ring of integers modulo $q$.  We denote elements of $\mathbb{Z}_q$ simply by $0, 1, 2, \ldots, q-1 $.  
For any $x \in \mathbb{Z}_q$, let $\left| x \right| \in \mathbb{Z}$ denote the minimum absolute value among all integers congruent to $x$ mod $q$ (e.g., in the ring $\mathbb{Z}_5$, $|1| = |4| = 1$ and $|2| = |3| = 2$).  For any vector $\mathbf{v} \in \mathbb{Z}_q^n$, let $\left\| \mathbf{v} \right\|_1 =  \sum_{j=1}^n \left| v_j \right|$, and let $\left\| \mathbf{v} \right\|_\infty$ denote the maximum value of $\left| v_j \right|$ among all coordinates $v_j$ of $\mathbf{v}$.

\begin{remark}
\label{rem:bin}
We use the big-endian convention for binary representations. 
 For any $x \in \mathbb{Z}_q$, let $[x] \in \{ 0, 1 \}^{\lceil \log z \rceil}$ denote the binary representation of $x$ in big-endian order (for example, if $q = 11$ and $x = 5$, then $[x] = 0101$).  If $\mathbf{x} \in \mathbb{Z}_q^n$, then $[\mathbf{x}]$ denotes the length-$(n \lceil \log z \rceil)$ binary sequence $[x_1][x_2]\ldots [x_n]$.  For any $j \in \{ 1, 2, \ldots, n \lceil \log z \rceil \}$, we denote by $[\mathbf{x}]_j$ the $j$th bit of $[\mathbf{x}]$.  If 
 $\mathbf{j} = (j_1, j_2, \ldots, j_k)$ is an increasing sequence drawn from $\{ 1, 2, \ldots, n \lceil \log z \rceil \}$, then $[ \mathbf{x}]_\mathbf{j}$ denotes the length-$k$ binary sequence
 \begin{eqnarray}
     [\mathbf{x}]_{j_1} [\mathbf{x}]_{j_2} \ldots [\mathbf{x}]_{j_k}.
 \end{eqnarray}
\end{remark}

A \textbf{register} is simply a finite set $\mathcal{Q}$ (the set of basic states).  A quantum register also has an associated complex Hilbert space $Q = \left\{ f \mid f \colon \mathcal{Q} \to \mathbb{C} \right\}$.  A \textbf{qubit}  is a quantum register $\mathcal{Q}$ with a fixed bijection $\mathcal{Q} \leftrightarrow\{ 0, 1 \}$.

Let $\mathcal{V}$ be a quantum register and $V$ its associated complex Hilbert space. We denote by $\mathbb{I}_V$ the identity operator on $V$.
A \textbf{positive operator-valued measure} (POVM) on $V$ is a finite set $\{ M_y \}_{y \in Y}$ of positive semidefinite operators on $V$ satisfying $\sum_y M_y = \mathbb{I}_V$.
A \textbf{pure state} on $V$ is a unit vector in $V$.  A \textbf{mixed state} on $V$ is a positive semi-definite linear operator on $V$ of trace $1$. A \textbf{subnormalized} mixed state is a positive semi-definite operator linear operator in $V$ of trace less than or equal to $1$.

If $m$ is a positive integer and $\textbf{w} = (w_1, w_2, \ldots, w_m ) \in \mathbb{C}^m$, then $\left\| \textbf{w} \right\|_1$ denotes the $1$-norm of $w$:
\begin{eqnarray}
\left\| \mathbf{w} \right\|_1 & = & \sum_{j=1}^m \left| w_j \right|.
\end{eqnarray}
The expression $\left\| \textbf{w} \right\|$ denotes the Frobenius norm:
\begin{eqnarray}
\left\| \mathbf{w} \right\| & = & \sqrt{ \sum_{j=1}^m \left| w_j \right|^2 }.
\end{eqnarray}
The \textbf{Cauchy-Schwarz inequality} is simply the obervation that if $W$ is a finite-dimensional complex Hilbert space and $\textbf{x}, \textbf{y} \in W$, then
\begin{eqnarray}
    \left| \left< \textbf{x}, \textbf{y} \right> \right| & \leq & \left\| \textbf{x} \right\| \left\| \textbf{y} \right\|,
\end{eqnarray}
with equality if and only if $\textbf{x}$ and $\textbf{y}$ are parallel vectors.

If $z$ is a complex number and $z = r e^{i \theta}$ with $0 \leq \theta < 2 \pi$ and $r \geq 0$, and if $c$ is a real number, then we define $z^c$ to mean the quantity
\begin{eqnarray}
    z^c & = & r^c e^{i \theta c}.
\end{eqnarray}
If $D$ is an $m \times m$ diagonal matrix with diagonal entries $d_1, d_2, \ldots, d_m \in \mathbb{C}^m$, then $D^c$ denotes the diagonal matrix with entries $d_1^c, d_2^c, \ldots, d_m^c$.

We write $\left| 0 \right>, \left| 1 \right>, \left| + \right>, \left| - \right>$ to denote, respectively, the following vectors on $\mathbb{C}^2$:
\begin{eqnarray}
    (1, 0), (0, 1), \left( \frac{1}{\sqrt{2}} , \frac{1}{\sqrt{2}} \right) , \left( \frac{1}{\sqrt{2}} , - \frac{1}{\sqrt{2}} \right).
\end{eqnarray}
(We may omit the Dirac brackets $\left| \cdot \right>$ when it is convenient to do so.)  The letters $X$, $Y$, $Z$ denote the Pauli operators:
\begin{eqnarray}
    \begin{array}{ccc}
    X = \left[ \begin{array}{cc} 0 & 1 \\ 1 & 0 \end{array} \right] 
    \hskip0.5in
    Y = \left[ \begin{array}{cc} 0 & i \\ -i & 0 \end{array} \right] 
    \hskip0.5in
    Z = \left[ \begin{array}{cc} 1 & 0 \\ 0 & -1 \end{array} \right]. 
    \end{array}
\end{eqnarray}
If $W$ is a Hermitian operator on $\mathbb{C}^2$ with eigenvalues in $\{ -1, +1 \}$, then measuring ``in the $W$-basis'' means applying the POVM $\{ (\mathbb{I} + W )/2, (\mathbb{I} - W )/2 \}$ to obtain a bit $w$ (where $(\mathbb{I} + W )/2$ corresponds to outcome $w = 0$ and $(\mathbb{I} - W )/2$ corresponds to outcome $w = 1$).

Let $S$ be a finite set and let $p \colon S \to [0, 1 ]$ be a probability distribution on $S$.  Then, $s \leftarrow p$ denotes that the element $s$ is sampled from $S$ according to $p$, and $s \leftarrow S$ denotes that $s$ is sampled from $S$ via a uniform distribution.
The \textbf{collision probability} of $p$ is the quantity
\begin{eqnarray}
    \sum_{s \in S} p_s^2.
\end{eqnarray}
(This quantity is equal to the probability that two independent samples from $p$ will agree.)  We note the following proposition, which follows directly from the Cauchy-Schwartz inequality.
\begin{proposition}
\label{prop:collision}
Let $S$ be a finite set, let $p, p' \in S$ be probability distributions on $S$, and let $c = \sum_{s\in S} p_s^2$ and $c' = \sum_{s \in S} (p'_s)^2$.  Then,
\begin{eqnarray}
    \mathbf{P} \left[ s = s' \mid s \leftarrow p, s' \leftarrow p' \right] & \leq & \sqrt{ cc'},
\end{eqnarray}
with equality if and only if $p=p'$. $\qed$

\end{proposition}

We note a few more details about notation and conventions:

\begin{itemize}
    \item 
If $S$ is a set and $f \colon S \to \mathbb{C}$ is a function, then $\Supp f$ (the support of $f$) is the set of all elements of $S$ on which $f$ takes a nonzero value.

\item If $S$ and $T$ are sets, then $S^T$ denotes the set of functions from $T$ to $S$.

\item
If $S$ is a finite set, then the vector space $\mathbb{C}^S$ has an inner product given by $\left< f , g \right> = \sum_{s \in S} f(s) \overline{g(s)}$.  Therefore, we consider $\mathbb{C}^S$ to be a finite-dimensional complex Hilbert space.

\item
If $Z$ is a logical expression, then $\delta_Z$ denotes an indicator value for $Z$: if $Z$ is false, then $\delta_Z = 0$, and if $Z$ is true, then $\delta_Z = 1$.
\end{itemize}

\subsection{Nonlocal Games}

\begin{definition}
\label{def:nongame}
    A \text{nonlocal $m$-player game} (for $m \geq 2$) consists of the following data:
    \begin{itemize}
        \item Finite sets $\mathcal{X}_1, \mathcal{X}_2, \ldots, \mathcal{X}_m$ and $\mathcal{A}_1, \mathcal{A}_2, \ldots, \mathcal{A}_m$.

        \item A probability distribution $r$ on $\mathcal{X}_1 \times \cdots \times \mathcal{X}_m$.

        \item A scoring function
        \begin{eqnarray}
            F \colon \mathcal{X}_1 \times \cdots \times \mathcal{X}_m \times \mathcal{A}_1 \times \cdots \times \mathcal{A}_m \to \mathbb{R}.
        \end{eqnarray}
    \end{itemize}
\end{definition}

Given the data above, a nonlocal $m$-player game proceeds as follows.  We refer to the participants in the game as Referee and Alice$_1$, Alice$_2$, \ldots, Alice$_m$.
\begin{enumerate}
    \item Referee samples a sequence $(x_1, \ldots, x_m)$ according to $r$ and sends $x_i$ to Alice$_i$. 

    \item Alice$_i$ outputs an element $a_i \in \mathcal{A}_i$.

    \item The Referee awards the score $F(x_1, \ldots, x_m, a_1, \ldots, a_m)$.
\end{enumerate}

We will also need the definition of a sequential nonlocal two-player game.  (See \cite{gallego2014nonlocality} for some related formalism.)

\begin{definition}
    \label{def:seqgame}
    A \text{sequential nonlocal $m$-player game} is a nonlocal game
    \begin{eqnarray}
        \left( \mathcal{X}_1, \ldots, \mathcal{X}_m, \mathcal{A}_1, \ldots, \mathcal{A}_m,  r, F \right)
    \end{eqnarray}
    with additional positive integer parameters $d_1, d_2, \ldots, d_m$ such that for every $i \in \{ 1, 2, \ldots, m \}$,
    \begin{eqnarray}
        \label{xdecomp} \mathcal{X}_i  & = & \mathcal{X}_{i,1} \times \mathcal{X}_{i,2} \times \cdots \times \mathcal{X}_{i, d_i} \\
        \label{adecomp} \mathcal{A}_i  & = & \mathcal{A}_{i,1} \times \mathcal{A}_{i,2} \times \cdots \times \mathcal{A}_{i, d_i}.
    \end{eqnarray}
\end{definition}

A sequential nonlocal two-player game proceeds as follows.
\begin{enumerate}
    \item Referee samples a sequence $(x_1, \ldots, x_m)$ according to $r$ and sends $x_i$ to Alice$_i$. 

    \item \label{iterativestep} For $j = 1, 2, \ldots d_1$, Referee sends the $j$th component of $x_1$ to Alice$_1$ and receives output $a_{1,i} \in \mathcal{A}_{i,1}$.  She records the outputs as $a_1 := (a_{1,1}, a_{1,2}, \ldots, a_{1,d_1})$.  

    \item \label{iterativestep2} Referee repeats Step~\ref{iterativestep} for each of Alice$_2$, Alice$_3$, \ldots, Alice$_m$ to obtain a sequence $(a_1, a_2, \ldots, a_m)$.

    \item The Referee awards the score $F(x_1, \ldots, x_m, a_1, \ldots, a_m)$.
\end{enumerate}
Crucially, in steps~\ref{iterativestep}--\ref{iterativestep2}, Alice$_i$ must return her output $a_{i,j}$ before receiving the next input $x_{i,j+1}$.  

\begin{definition}
    \label{def:qstrat}
    Let $H = \left( \mathcal{X}_1, \ldots, \mathcal{X}_m, \mathcal{A}_1, \ldots, \mathcal{A}_m,  r, F \right)$ be an $m$-player nonlocal game.  A \textbf{quantum strategy} for $H$ consists of the following data.
    \begin{itemize}
        \item Quantum registers $A_1, A_2, \ldots, A_m$

        \item A pure state $\psi \in A_1 \otimes \cdots \otimes A_m$,

        \item For every $i \in \{ 1, 2, \ldots, m \}$ and every $x \in \mathcal{X}_i$, a POVM $\left\{ M_{i,x}^a \mid a \in \mathcal{A}_i \right\}$ on $A_i$.
    \end{itemize}
\end{definition}

These data specify how Alice$_1$, $\ldots$, Alice$_m$ act when playing Game $G$:
\begin{enumerate}
    \item Before receiving their inputs, Alice$_1$, $\ldots$, Alice$_m$ share state $\psi$ (with Alice$_i$ possessing register $A_i$).

    \item When Alice$_i$ receives her input $x_i$ she applies the POVM $\{ M_{i,x}^a \}$ and outputs the result.
\end{enumerate}

We can compute the probabilistic behavior of the players directly in terms of the mathematical objects defined above: for example, if $m=2$, the probability that Alice and Bob will output $(a_1, a_2)$ on
input $(x_1,x_2)$ is given by
\begin{eqnarray}
\psi^* ( M_{1,x_1}^{a_1} \otimes M_{2,x_2}^{a_2 } ) \psi.
\end{eqnarray}
We refer to the subnormalized states
\begin{eqnarray}
    \Tr_{A_2} \left[ \left( \mathbb{I}_{A_1} \otimes M^{a_2}_{2,x_2} \right) \psi \psi^* \right]
\end{eqnarray}
for $a_2 \in \mathcal{A}_2, x_2 \in \mathcal{X}_2$ as the \textbf{steered states} for the first player, and the subnormalized states
\begin{eqnarray}
    \Tr_{A_1} \left[ \left(  M^{a_1}_{1,x_1} \otimes \mathbb{I}_{A_2} \right) \psi \psi^* \right]
\end{eqnarray}
as the steered states for the second player.  

\begin{definition}
        Let $H$ be a nonlocal $m$-player game (with notation as in Definition~\ref{def:nongame}) .  A \textbf{deterministic strategy} for $H$ is an $m$-tuple $(S_1, S_2, \ldots, S_m)$ of functions
    $S_i \colon \mathcal{X}_i \to \mathcal{A}_i$.  A \textbf{randomized strategy} for $H$  is a probability distribution on the set of all  deterministic strategies for $H$.
\end{definition}

\begin{definition}
    \label{def:time}
    Let $d$ be a positive integer and let $Y_1, Y_2, \ldots, Y_d, Z_1, Z_2, \ldots, Z_d$ be finite sets.  Then a function
    \begin{eqnarray}
        f \colon Y_1 \times \cdots \times Y_d \to Z_1 \times \cdots \times Z_d
    \end{eqnarray}
    is \textbf{time-ordered} if for all $i \in \{ 1, 2, \ldots, d-1 \}$, the $i$th component of $f ( y_1, y_2, \ldots, y_d )$ is independent of the values $y_{i+1}, y_{i+2}, \ldots y_d$.

    Let $G$ be a sequential nonlocal $m$-player game (with notation as in Definition~\ref{def:seqgame}).  Then, a deterministic strategy for $G$ is an $m$-tuple $(S_1, S_2, \ldots, S_m)$ of functions
    $S_i \colon \mathcal{X}_i \to \mathcal{A}_i$ such that for all $i \in \{ 1, 2, \ldots, m \}$, $S_i$ is time-ordered with respect to equations (\ref{xdecomp}) and (\ref{adecomp}).
\end{definition}

If $H$ is a game and $S$ is a strategy, then we will write $\omega ( H, S)$ for the expected score of $S$.  (For convenience, we will use this same notation regardless of whether $H$ is sequential or whether $S$ is a deterministic or quantum strategy.)

\begin{definition}
    Let $H$ be an $m$-player nonlocal game.   Then, \begin{itemize}
        \item The \textbf{classical value} of $H$, denoted $\omega^c ( H )$, is the supremum of $\omega ( H, S )$ over all deterministic strategies $S$.  The \textbf{classical bias} of $H$, denoted $\beta^c ( H )$, is the supremum of $\left| \omega ( H , S ) \right|$ over all deterministic strategies $S$.  \item  The \textbf{entangled value} of $H$, denoted $\omega^q ( H )$, is the supremum of $\omega ( H, T )$ over all quantum strategies $T$.  The \textbf{quantum bias} of $H$, denoted $\beta^q ( H )$, is the supremum of $\left| \omega ( H , T ) \right|$ over all quantum strategies $T$.  
        \end{itemize}
\end{definition}

\begin{definition}
    Let $H$ be an $m$-player nonlocal game.  The \textbf{quantum-to-classical bias ratio} of $G$ is the quantity
    \begin{eqnarray}
        \frac{\beta_q ( H )}{\beta_c ( H )}.
    \end{eqnarray}
\end{definition}

Let $H$ be an $m$-player nonlocal game whose scoring function has range $\{ 0, 1 \}$, and let $k$ be a positive integer. Then $H^d$ denotes $n$-fold parallel repeated game in which Alice$_1$, $\ldots$, Alice$_m$ play $d$ instances of game $H$ simultaneously.  The score in $H^d$ is $1$ if Alice$_1$, $\ldots$, Alice$_m$ win all $d$ of the games, and $0$ if any of the $n$ games is lost.  Let $H^{[d]}$ denote the sequential nonlocal game that arises from having the players play $d$-rounds of game $H$ in sequence.  While it is easy to show that
\begin{eqnarray}
    \omega^c ( H^{[d]} ) & = & \omega^c ( H )^d.
\end{eqnarray}
the same inequality does not necessarily hold when $H^{[d]}$ is replaced by $H^d$ \cite{raz1995parallel}.

\subsection{The Fourier Transform for Finite Abelian Groups}

We fix some conventions and terminology for the Fourier transform over finite abelian groups.  
If $G$ is a finite abelian group, then $\hat{G}$ (the dual group of $G$) denotes the set of all functions
\begin{eqnarray}
    h \colon G \to \mathbb{C} \smallsetminus \{ 0 \}
\end{eqnarray}
that satisfy $h(x+y) = h(x)h(y)$ for all $x,y \in G$. 

\begin{definition}
\label{def:fourier}
Let $G$ be a finite abelian group and let $f \colon G \to \mathbb{C}$ be a function.  Then, the Fourier transform of $f$, denoted $\hat{f}$, is the function from $\hat{G}$ to $\mathbb{C}$ given by
\begin{eqnarray}
    \hat{f} ( h ) & = & \left| G \right|^{-1/2} \sum_{x \in G } 
    f(x) \overline{h(x)}
\end{eqnarray}
\end{definition}
 Note that in the case where $G = \mathbb{Z}_m^n$ and $m \geq 2$ and $n$ are positive integers, there is a natural isomorphism $i \colon G \to \hat{G}$ given by 
\begin{eqnarray}
i(x)(x') & = & \zeta^{- x \cdot x'}
\end{eqnarray}
for all $x,x' \in G$,
where $\zeta = \exp ( 2 \pi i / m)$ and $x \cdot x'$ denotes the dot product between $x$ and $x'$.  In that case, we can consider the Fourier transformed function $\hat{f}$ from Definition~\ref{def:fourier} as a function on $G$ itself given by
\begin{eqnarray}
    \hat{f} ( x' ) & = & \left| G \right|^{-1/2} \sum_{x \in G } 
    f(x) \zeta^{x \cdot x'}.
\end{eqnarray}

\begin{definition}
    Let $G$ be a finite abelian group.  If $f,g \in \mathbb{C}^G$, then the function $f * g \in \mathbb{C}^G$ is defined by the expression
    \begin{eqnarray}
    (f*g)(x) & = &  \left| G \right|^{-1/2} \sum_{y \in G} f(x-y) g(y).
    \end{eqnarray}
\end{definition}
(The $*$ operation is commonly called convolution, although some authors normalize it differently.)  The following facts are easy to prove:
\begin{itemize}
\item The groups $G$ and $\hat{G}$ are always of the same cardinality.

\item If $f \in \mathbb{C}^G$, then
\begin{eqnarray}
    \left\| \hat{f} \right\|_2 & = & \left\| f \right\|_2.
\end{eqnarray}

\item If $f, g \in \mathbb{C}^G$, then
\begin{eqnarray}
    \widehat{f * g} & = & \hat{f} \cdot \hat{g},
\end{eqnarray}
where $\hat{f} \cdot \hat{g}$ is the function on $\hat{G}$ defined by $(\hat{f} \cdot \hat{g})(h) = \hat{f}(h) \hat{g} ( h )$.
\end{itemize}

\section{An Uncertainty Principle for Functions on Finite Abelian Groups}

\label{sec:uncertainty}

Let $r \colon \mathbb{R} \to \mathbb{C}$ be an infinitely differentiable function such that the Frobenius norm
\begin{eqnarray}
    \left\| r \right\|_2 & = & \sqrt{ \int_{-\infty}^{\infty} \left| r ( t ) \right|^2 dt} 
\end{eqnarray}
is equal to $1$. Let $\hat{r} \colon \mathbb{R} \to \mathbb{C}$ denote the Fourier transform
\begin{eqnarray}
    \hat{r} ( x ) & = & \int_{-\infty}^{\infty} r ( t ) e^{-2 \pi i xt}
\end{eqnarray}
Then, the Heisenberg uncertainty principle, interpreted mathematically (see Theorem~4.9 in \cite{wigderson2021uncertainty}), asserts that the product of the variances 
\begin{eqnarray}
    Var ( r ) & = & 
     \left( \int_{-\infty}^{\infty} t^2 \left| r ( t ) \right|^2 dt \right) - 
    \left( \int_{-\infty}^{\infty} t \left| r ( t ) \right|^2 dt \right)^2  \\
    Var ( \hat{r} ) & = & 
     \left( \int_{-\infty}^{\infty} t^2 \left| \hat{r} ( t ) \right|^2 dt \right) - 
    \left( \int_{-\infty}^{\infty} t \left| \hat{r} ( t ) \right|^2 dt \right)^2  
\end{eqnarray}
is bounded below by a constant. 

It is natural to ask whether a similar assertion exists for the Fourier transform over finite abelian groups.  A commonly used uncertainty principle in the finite abelian case, attributed to Donoho and Stark \cite{donoho1989uncertainty}, is the following.  Rather than measuring the uncertainty of a function in terms of its variance, one measures it in terms of the size of the support of the function.  A simple proof can be found in \cite{wigderson2021uncertainty} (see subsection 3.1).
\begin{theorem}[\cite{donoho1989uncertainty}]
    \label{thm:donoho}
    Let $G$ be a finite abelian group, and let $f \colon G \to \mathbb{C}$ be a nonzero function.  Then,
    \begin{eqnarray}
        \label{donohoineq}
        \left| \Supp f \right| \left| \Supp \hat{f} \right| & \geq & \left| G  \right|. \qed
    \end{eqnarray}
\end{theorem}
A number of variants of Theorem~\ref{thm:donoho} are stated in section~3 of \cite{wigderson2021uncertainty}.
For the purpose of this work, we need a theorem that addresses the possible configurations of the sets $\Supp f$ and $\Supp \hat{f}$  that can occur when inequality (\ref{donohoineq}) is close to an equality.  This topic is touched on in papers such as \cite{donoho1989uncertainty, tao2005uncertainty}, but since we need a quantifiable assertion, we will prove a different variant of Theorem~\ref{thm:donoho} from scratch.

We begin with two definitions.

\begin{definition}
    Let $G$ be a finite abelian group, and let $f \colon G \to \mathbb{C}$ be a nonzero function.
    Let $p$ be the probability distribution defined by $p = \left| f \right| / \left\| f \right\|_1$.
    Then, 
    the \textbf{uniformity coefficient} of $f$
    is the quantity
    \begin{eqnarray}
        \nu ( f ) & = & \frac{\mathbf{P}
        [x = y \mid x, y \leftarrow \Supp p ]}{\mathbf{P}
        [x = y \mid x,y \leftarrow p ]} \\
        & = & \frac{1}{\left| \Supp p \right| \mathbf{P}
        [x = y \mid x, y\leftarrow p ]}.
    \end{eqnarray}
 \end{definition}

\begin{definition}
    \label{def:linearity}
    Let $G$ be a finite abelian group, and let $f \colon G \to \mathbb{C}$ be a nonzero function.
    Let $p$ be the probability distribution defined by $p = \left| f \right| / \left\| f \right\|_1$.
    Then, 
    the \textbf{linearity coefficient} of $f$
    is the quantity
    \begin{eqnarray}
        \eta ( f ) & = & \frac{\mathbf{P}
        [x + y = z + w \mid x, y,z,w \leftarrow p ]}{\mathbf{P}
        [x = y \mid x,y\leftarrow p ]}.
    \label{lindef}
    \end{eqnarray}
    If $S$ is a nonempty subset of $G$, then the linearity coefficient of $S$, denoted $\eta ( S )$, is the linearity coefficient of the indicator function $g(x) = \delta_{x \in S}$.
 \end{definition}

\begin{proposition}
    \label{prop:unif}
    Let $G$ be a finite abelian group and $f \colon  G \to \mathbb{C}$ a nonzero function.  Then, $\nu ( f ) \leq 1$, with equality if and only if $|f|$ is constant on $\Supp f$.
\end{proposition}

\begin{proof}
    Let $p = \left| f \right| / \left\| f \right\|_1$. 
    By the Cauchy-Schwartz inequality, we have
    \begin{eqnarray}
    \nu ( f )^{-1} & = & \left( \sum_{x \in \Supp p} p(x)^2 \right)
    \left( \sum_{x \in \Supp p} 1 \right) \\
    & \geq & \left( \sum_{x \in \Supp p} p ( x ) \cdot 1 \right)^2 \\
    & = & 1,
    \end{eqnarray}
with equality if and only $p$ is constant on $\Supp p$.   This completes the proof.  
\end{proof}

\begin{proposition}
    \label{prop:lincoeff}
    Let $G$ be a finite abelian group and let $f \colon G \to \mathbb{R}_{\geq 0}$ be a nonzero function.  Then, $\eta ( f ) \leq 1$.  Moreover, if $\eta ( f ) = 1$, then $\Supp f$ must be a coset of a subgroup of $G$.
\end{proposition}

\begin{proof}
Let $\alpha$ denote the collision probability of $p$.  By Proposition~\ref{prop:collision}, we have 
\begin{eqnarray}
    \mathbf{P} \left[ x + y = z + w \mid x,y,z,w \leftarrow p \right] 
    & = & \sum_{y,w \in \Supp p} p(y) p ( w ) \mathbf{P} [ x + y = z + w \mid x , z \leftarrow p] \\
    & \leq & \sum_{y,w \in \Supp p} p ( y ) p ( w )  \sqrt{\alpha^2 } \label{alphaline} \\
    & = & \alpha.
\end{eqnarray}
We conclude that $\eta ( f ) \leq 1$.  

Now suppose that $\eta ( f ) = 1$.  Then, equality occurs in line (\ref{alphaline}) above and therefore (also by Proposition~\ref{prop:collision}) for any $y, w \in \Supp p$, the probability distributions $[ x  + y \mid x \leftarrow p]$ and $[z + w \mid z \leftarrow p ]$ must be the same.  In particular, this means that these two distributions have the same support. Therefore $\Supp p$ satisfies the following condition:
\begin{eqnarray}
x,y,z \in \Supp p & \Longrightarrow & x + y - z \in \Supp p.
\end{eqnarray}
Letting $H = (-y) + \Supp p$, we have that for any $x,z \in \Supp p$, $(x-y) - (z-y) = x-z \in -y + \Supp p = H$. We
conclude that $H$ is closed under differences:
\begin{eqnarray}
a ,b \in  H & \Longrightarrow & a - b \in H.
\end{eqnarray}
For any $b \in H$, the subtraction map $x \mapsto x - b$ on $G$ is a permutation on $G$ that maps $H$ to itself, and therefore its inverse $x \mapsto x + b$ also maps $H$ to itself.  We conclude that $H$ is also closed under addition.   For any $a \in \Supp p$, since $G$ is finite, we must have $n \cdot a = 0$ for some $n$, and therefore $0 = n \cdot a \in H$ and $-a = (n-1) \cdot a \in H$.  We conclude that $H$ is a subgroup of $G$.  Therefore, $\Supp p = y + H$ is a coset of a subgroup of $G$.
\end{proof}

The following is the main result of this section.

\begin{theorem}
\label{thm:uncertainty}
Let  $G$ be a finite abelian group, and let $f \colon G \to \mathbb{C}$ and $g \colon \hat{G} \to \mathbb{C}$ be functions such that $\left\| g \right\|_2 = \left\| f \right\|_2 = 1$.  
Then,
\begin{eqnarray}
    \left| \left< \hat{f}, g \right> \right| & \leq & \left( \frac{ \left| \Supp f \right| \left| \Supp g \right| \nu ( f) \eta ( f ) }{\left| G \right| } \right)^{1/4}.
\end{eqnarray}
\end{theorem}

Before presenting the proof of Theorem~\ref{thm:uncertainty},  we note the following corollary.
\begin{corollary}
\label{cor:uncertainty}
Let $G$ be a finite abelian group and let $h \colon G \to \mathbb{C}$ be a nonzero function.  Then,
\begin{eqnarray}
    \frac{ \left| \Supp h \right| \left| \Supp \hat{h} \right| \nu ( h ) \eta ( h )  }{\left| G \right| }  & \geq & 1.
\end{eqnarray}
\end{corollary}
\begin{proof}
    Apply Theorem~\ref{thm:uncertainty} with $f =  h/ \left\| h \right\|_2$ and $g = \hat{f}$.
\end{proof}
This corollary implies that if the ratio of $\left| \Supp h \right| \left| \Supp \hat{h} \right|$ to $\left| G \right|$ is close to $1$, then $\nu ( h )$ and $\eta ( h )$ must be close to $1$. Thus, informally, if equality is nearly achieved in Theorem~\ref{thm:donoho}, then $|h|$ is close to being uniformly supported on a coset of a subgroup of $G$.

\begin{proof}[Proof of Theorem~\ref{thm:uncertainty}]
We begin by applying the Cauchy-Schwartz inequality twice to get an upper bound on $\left| \left< \hat{f} , g \right> \right|$:
\begin{eqnarray}
    \left| \left< \hat{f} , g \right> \right|^4 & = &
    \left| \sum_{x \in \Supp g} \hat{f} ( x ) \overline{g ( x )} \right|^4 \\
    & \leq & \left( \left( \sum_{x \in \Supp g } \left| \hat{f} ( x ) \right|^2 \right)
    \left( \sum_{x \in \Supp g} \left| g ( x ) \right|^2 \right) \right)^2 \\
    & = & \left( \left(  \sum_{x \in \hat{G}} \left| \hat{f} ( x ) \right|^2 \cdot \delta_{x \in \Supp g} \right) (1 ) \right)^2 \\
    & \leq & \left( \sum_{x \in \hat{G}} \left| \hat{f} ( x ) \right|^4 \right) \left( \sum_{x \in \hat{G}} \delta_{x \in \Supp g}^2 \right) \\
    \label{fourthpower}
    & = & \left( \sum_{x \in \hat{G}} \left| \hat{f} ( x ) \right|^4 \right) \left| \Supp g \right| 
\end{eqnarray}
Observing that the first factor in the product (\ref{fourthpower}) is equal to $\left\| \hat{f} \cdot \hat{f} \right\|_2^2$, where $\hat{f} \cdot \hat{f}$ denotes the pointwise product of $\hat{f}$ with itself, we have
\begin{eqnarray}
    \left| \left< \hat{f} , g \right> \right|^4    & \leq &  \left\| \hat{f} \cdot \hat{f} \right\|_2^2 \left| \Supp g \right| \\
    & = &  \left\| \widehat{f * f} \right\|_2^2  \left| \Supp g \right| \\
    & = & \left\| f * f \right\|_2^2  \left| \Supp g \right| \\
    & = & \left( \sum_{x \in G} (f * f) (x) \overline{(f * f ) (x)}  \right)  \left| \Supp g \right| \\
    & = & \left| G \right|^{-1} \left( \sum_{x \in G} \left( \sum_{\substack{ y,z \in G \\ y + z = x } }  f(y) f(z) 
    \right) \left( \sum_{\substack{ y',z' \in G \\
    y' + z' = x} } \overline{f(y')} \overline{f(z')} \right) \right) \left| \Supp g \right| \\
    & = & \left| G \right|^{-1} \left(  \sum_{\substack{ y,z,y',z' \in G \\ y + z = 
    y' + z'} } f(y) f(z) \overline{f(y')} \overline{f(z')} \right) \left| \Supp g \right| \\
    & \leq & \left| G \right|^{-1} \left(  \sum_{\substack{ y,z,y',z' \in G \\ y + z = 
    y' + z'} } \left| f(y) f(z) f(y') f(z') \right| \right) \left| \Supp g \right| 
\end{eqnarray}
Let $p = \left| f \right| / \left\| f \right\|_1$, and let 
$\alpha = \mathbf{P} \left[ x = y \mid x,y \leftarrow p \right]$ denote the collision probability for $p$.
Note that
\begin{eqnarray}
\alpha & = &
\sum_{x \in G} p(x)^2 \\
& = & \left\| f \right\|_1^{-2} \sum_{x \in G}  \left| f ( x ) \right|^2 \\
& = & \left\| f \right\|_1^{-2}.
\end{eqnarray}
We have
\begin{eqnarray}
    \left| \left< \hat{f} , g \right> \right|^4  
    & \leq &  \left| G \right|^{-1} \left\| f \right\|_1^4 \left(  \sum_{\substack{ y,z,y',z' \in G \\ y + z = 
    y' + z'} }  p(y) p(z) p(y') p(z')  \right) \left| \Supp g \right| \\
    & = &  \left| G \right|^{-1} \alpha^{-2} \left(  \sum_{\substack{ y,z,y',z' \in G \\ y + z = 
    y' + z'} }  p(y) p(z) p(y') p(z')  \right) \left| \Supp g \right| \\
    & = &  \left| G \right|^{-1} \alpha^{-2} \left( \mathbf{P} \left[ y+z = y'+z' \mid y,z,y',z' \leftarrow p \right] \right) \left| \Supp g \right| \\
    & = &  \left| G \right|^{-1} \alpha^{-1} \eta ( f )  \left| \Supp g \right| \\
    & = & \left| G \right|^{-1} \left| \Supp f \right| \nu ( f ) \eta (f ) \left| \Supp g \right|,
\end{eqnarray}
which implies the desired result.
\end{proof}

\section{The GHZ Games}

The GHZ games \cite{mermin1990extreme} are binary games played by $3$ or more players.

\begin{definition}
\label{def:ghz}
    Let $k \geq 3$ be an integer.  Then, the $k$-player game $GHZ_k$ is defined as following:
\begin{itemize}
    \item The input and output alphabets for each player are $\{ 0, 1 \}$.
    
    \item The input probability distribution is given as follows,
    for $\mathbf{x} \in \{ 0, 1 \}^k$:
    \begin{eqnarray}
        r ( \mathbf{x} ) & = & \left\{ \begin{array}{cl} 2^{-k+1} & \textnormal{ if } x_1 \oplus \cdots \oplus x_k = 0 \\ \\
        0 & \textnormal{ otherwise.} \end{array} \right.
    \end{eqnarray}

    \item The score is equal to $1$ if 
    \begin{eqnarray}
          x_1 + x_2 + \ldots + x_k + 2 a_1 + 2 a_2 + \ldots + 2 a_k  & \equiv & 0 \hskip0.1in (\textnormal{mod } 4 ).
    \end{eqnarray}
    and is equal to $0$ otherwise.
\end{itemize}
\end{definition}

The entangled value of $GHZ_k$ is always $1$, while the classical value can be shown to decrease exponentially in $k$ by elementary arguments (see \cite{mermin1990extreme}).  A considerably more difficult problem is to upper bound the classical value of the parallel repeated game $GHZ_k^d$ when $k$ is fixed and $d$ tends to infinity.  The following result was recently proved.
\begin{theorem}[Theorem I.1 in \cite{braverman2023parallel}]
The parallel repeated games $GHZ_3^d$ satisfy
\begin{eqnarray}
    \omega^c ( GHZ_3^d ) & \leq & \exp ( - \Omega ( d )). \qed
\end{eqnarray}
\end{theorem}

The following corollary is a consequence.

\begin{corollary}
\label{cor:fourghz}
The parallel repeated games $GHZ_4^d$  satisfy
\begin{eqnarray}
    \omega^c ( GHZ_4^d ) & \leq & \exp ( - \Omega ( d )).
\end{eqnarray}
\end{corollary}

\begin{proof}
    See Appendix~\ref{app:fourghz}.
\end{proof}

\subsection{Parity-balanced subsets of $\mathbb{Z}_4^d$}

\begin{definition}
    \label{def:parb}
    Let $d$ be a positive integer.  A set $S \subseteq \mathbb{Z}_4^d$ is a \textbf{parity-balanced subset} of $\mathbb{Z}_4^d$ if the quotient map
    \begin{eqnarray}
        \mathbb{Z}_4^d & \to & \mathbb{Z}_2^d
    \end{eqnarray}
    induces a bijection between $S$ and $\mathbb{Z}_2^d$.
\end{definition}

Informally, a subset of $\mathbb{Z}_4^d$ is parity-balanced if it is of size $2^d$ and if its residues mod $2$ are evenly distributed over $\mathbb{Z}_2^d$.  A parity-balanced set $S \subseteq \mathbb{Z}_4^d$ naturally gives a deterministic strategy for a single player at the $d$-fold repeated GHZ game: for an input bit string $\mathbf{x}$, there is a unique element $s_\mathbf{x} \in S$ that is congruent mod $2$ to $\mathbf{x}$, and the player returns the unique bit string $\mathbf{a}$ that satisfies
\begin{eqnarray}
    s_\mathbf{x} & = & \mathbf{x} + 2 \mathbf{a}.
\end{eqnarray}
To specify a full deterministic strategy $GHZ_k^d$, it suffices to specify a parity-balanced set $S_j \subseteq \mathbb{Z}_4^d$ for the $j$th player, for $j \in \{ 1, 2, \ldots, k \}$.  The expected winning probability of the corresponding strategy is then
\begin{eqnarray}
    && \mathbf{P} \left[ s_1 + s_2 + \ldots + s_k = 0 \mid (s_1, s_2, \ldots, s_k ) \leftarrow (S_1, S_2, \ldots, S_k ), s_1 + s_2 + \ldots + s_k \in 2 \mathbb{Z}_4^d \right] \\
    & = & 2^d \cdot \mathbf{P} \left[ s_1 + s_2 + \ldots + s_k = 0 \mid (s_1, s_2, \ldots, s_k ) \leftarrow (S_1, S_2, \ldots, S_k ) \right].
\end{eqnarray}
Let $T$ be a parity-balanced subset of $\mathbb{Z}_4^d$.  
If we let $k = 4$ and $S_1 = S_2 = T$, $S_3 = S_4 = -T$, the winning probability for the associated $GHZ_4^d$-strategy is
\begin{eqnarray}
    2^d \cdot \mathbf{P} \left[ t_1 + t_2 - t_3 - t_4 = 0 \mid t_1, t_2, t_3, t_4 \leftarrow T \right],
\end{eqnarray}
which is precisely equal to the linearity coefficient $\eta ( T )$ of $T$.  We therefore obtain the following proposition, which will be important in section~\ref{sec:clawgame}.
\begin{proposition}
\label{prop:parity}
    Let $T$ be a parity-balanced subset of $\mathbb{Z}_4^d$.  Then,
    \begin{eqnarray}
        \eta ( T) & \leq & \omega^c ( GHZ_4^d ).
    \end{eqnarray}
\end{proposition}
As a consequence, by Corollary~\ref{cor:fourghz}, any parity-balanced subset of $\mathbb{Z}_4^d$ has linearity coefficient that is upper-bounded by a uniform exponentially-vanishing function of $d$.

In analogy to Definition~\ref{def:time}, we make the following definition
\begin{definition}
    \label{def:timeparb}
    Let $d$ be a positive integer.  A set $S \subseteq \mathbb{Z}_4^d$ is a \textbf{time-ordered parity-balanced subset} of $\mathbb{Z}_4^d$ if it  is of the form
    \begin{eqnarray}
        S & = & \left\{ \mathbf{x} + 2 f ( \mathbf{x} ) \mid \mathbf{x} \in \{ 0 , 1 \}^d \right\}
    \end{eqnarray}
    where $f \colon \{ 0, 1 \}^d \to \{ 0, 1 \}^d$ is a time-ordered function.
\end{definition}

By the same construction as above, any time-ordered parity balanced subset $T \subseteq \mathbb{Z}_4^d$ yields a deterministic strategy for the sequential game $GHZ_4^{[d]}$ whose expected score is $\eta ( T )$.  Since $\omega^c ( GHZ^{[d]}_4 ) = \omega^c ( GHZ_4 )^d = (3/4)^d $ (see Appendix~\ref{app:ghzscore}), we have the following.

\begin{proposition}
    \label{prop:timeparity}
    Let $T$ be a time-ordered parity-balanced subset of $\mathbb{Z}_4^d$.  Then,
    \begin{eqnarray}
        \eta ( T ) & \leq & (3/4)^d.
    \end{eqnarray}
\end{proposition}

\section{A Nonlocal Game for Claw-States}

\label{sec:clawgame}

Our goal in this section is to construct a family of two-player nonlocal games, indexed by a positive integer $d$, satisfying the following properties:
\begin{itemize}
    \item The quantum-to-classical bias ratio is lower bounded by $\exp ( \Omega ( d ))$.

    \item For each $d$, there is an optimal strategy for $G_d$ in which the steered states for one of the players are claw-states of length $O ( d )$.

    \item The input and output alphabets are all of size $\exp ( O ( d ))$.
\end{itemize}

Game $\mathbf{J}_d$, in Figure~\ref{fig:gamej}, will be shown to satisfy all three of the above conditions.  It can be seen as a derivative of the Hidden Matching game \cite{buhrman2011near} and the GHZ games (Definition~\ref{def:ghz}).  We additionally define Game $\mathbf{J}'_d$ (Figure~\ref{fig:gamejprime}), which is a sequential version of $\mathbf{J}_d$ in which the communication between the referee and the second player (Bob) is done in $(d+1)$ rounds.

\begin{figure}
    \centering
    \fbox{\parbox{5.5in}{\textbf{Game} $\mathbf{J}_d$: \\

\textit{Parameter:} A positive integer $d$ \\
\textit{Parties:} Referee, Alice, Bob.

\begin{enumerate}
    \item Referee samples two bit strings $\mathbf{x}, \mathbf{y} \in \{ 0, 1 \}^{d+1}$ uniformly at random under the constraint $x_{d+1} = y_{d+1} = 1$.
    
    \item Referee sends $\mathbf{x}$ to Alice and receives output $\mathbf{a} \in \{ 0, 1 \}^{d+1}$.
    
    \item Referee sends $\mathbf{y}$ to Bob and receives output $\mathbf{b} \in \{ 0 , 1 \}^{d+1}$.

    \item Referee computes $\mathbf{u}, \mathbf{v} \in \mathbb{Z}^{d+1}$ as follows:
    \begin{eqnarray*}
        u_i & = & x_i (-1)^{a_i}, \\
        v_i & = & y_i + 2 b_i.
    \end{eqnarray*}

    \item Referee assigns a score of $+1$ if $\mathbf{u} \cdot \mathbf{v}$ is congruent to $0$ or $1$ mod $4$, and assigns a score of $-1$ if $\mathbf{u} \cdot \mathbf{v}$ is congruent to $2$ or $3$ mod $4$.
\end{enumerate}
    }}
    \caption{The nonlocal game $\mathbf{J}_d$.}
    \label{fig:gamej}
\end{figure}

\begin{figure}
    \centering
    \fbox{\parbox{5.5in}{\textbf{Game} $\mathbf{J}'_d$: \\

Game $\mathbf{J}'_d$ is the same as Game $\mathbf{J}_d$, except that the following sequential communication step replaces Step 3:

\begin{enumerate}
    \item[3.] For $i = 1, 2, 3, \ldots, d+1$, Referee sends $y_i$ to Bob and receives output $b_i \in \{ 0, 1\}^{d}$. 
\end{enumerate}
    }}
    \caption{The nonlocal game $\mathbf{J}'_d$.}
    \label{fig:gamejprime}
\end{figure}

\subsection{A Quantum Strategy for $\mathbf{J}_d$}

\begin{proposition}
    \label{prop:jent}
    Let $d$ be a positive integer.  There exists a quantum strategy for $\mathbf{J}_d$ that achieves an expected score of $\sqrt{2}/2$.  
\end{proposition}

\begin{proof}
    Let $A$ and $B$ be $(d+1)$-qubit registers, with $A = A_1 \otimes \ldots \otimes A_{d+1}$, $B = B_1 \otimes \ldots \otimes  B_{d+1}$ and $A_i \cong B_i \cong \mathbb{C}^2$ for all $i \in \{ 1, 2, \ldots, d+1 \}$.  Let $\psi \in \mathcal{A} \otimes B$ be the maximally entangled state
    \begin{eqnarray}
        \psi & = & 2^{-(d+1)/2} \sum_{\mathbf{z} \in \{ 0, 1 \}^{d+1} }\left| \mathbf{z} \right> \otimes \left| \mathbf{z} \right>.
    \end{eqnarray}
    For any $\mathbf{x}, \mathbf{a} \in \{ 0, 1 \}^{d+1}$ with $x_{d+1} = 1$, let $M_\mathbf{x}^\mathbf{a}$ denote the projector in $A$ onto the subspace spanned by 
    \begin{eqnarray}
        \label{theta}
        \theta_{\mathbf{x}}^\mathbf{a} & = & \frac{1}{\sqrt{2}} \left( \left| a_1, a_2 ,  \ldots, a_d , 0 \right> + (-1)^{a_{d+1}} \left|  a_1 \oplus x_1, a_2 \oplus x_2, \ldots, a_d \oplus x_d, 1 \right> \right)
    \end{eqnarray}
    For any $\mathbf{y}, \mathbf{b} \in \{ 0, 1 \}^{d+1}$ with $y_{d+1} = 1$, let
    $N_\mathbf{y}^\mathbf{b}$ denote the projector in $B$ onto the state
    \begin{eqnarray}
        \phi_{\mathbf{y}}^\mathbf{b} & = &
        \left( \bigotimes_{j=1}^d 
        \left( Z^{y_i/2} Z^{b_i} \left| + \right> \right) \right)
        \otimes\left( Z^{1/4} Z^{b_{d+1}} \left| + \right> \right)
    \end{eqnarray}
    We note (following previous work, e.g., \cite{alnawakhtha2024lattice}) that for any $j \in \{ 1, 2, 3, \ldots, d \}$, applying a gate of the form $Z^c$ (with $c \in \mathbb{R}$) to the $j$th qubit  of $\theta_\mathbf{x}^\mathbf{a}$ yields the same result as applying the gate
    \begin{eqnarray}
        Z^{x_j (-1)^{a_j} c}
    \end{eqnarray}
    to the $(d+1)$st qubit  of $\theta_{\mathbf{x}}^\mathbf{a}$.   
    Therefore, if we let $U \colon ( \mathbb{C}^2 )^{\otimes (d+1) } \to ( \mathbb{C}^2 )^{\otimes (d+1)}$ be the following unitary operator (which maps $\left| + \right>^{\otimes (d+1)}$ to $\phi^{\mathbf{b}}_\mathbf{y}$):
    \begin{eqnarray}
        U & = &  
        Z^{y_1/2} Z^{b_1} 
        \otimes  Z^{y_2/2} Z^{b_2} 
        \otimes \cdots \otimes
        Z^{y_d/2} Z^{b_d} \otimes Z^{1/4} Z^{b_{d+1}} ,
    \end{eqnarray}
    then, 
    \begin{eqnarray}
        \psi^* \left(  M_\mathbf{x}^\mathbf{a} \otimes N_\mathbf{y}^\mathbf{b}  \right) \psi & = & 2^{-(d+1)} \left| \left< \theta_{\mathbf{a}}^\mathbf{x} \mid \overline{ \phi_\mathbf{y}^\mathbf{b}}  \right> \right|^2 \\
& = & 2^{-(d+1)} \left| \left< \theta_{\mathbf{a}}^\mathbf{x}  \mid  U^{-1} (\mathbf{+}^{\otimes d} )  \right> \right|^2 \\
& = & 2^{-(d+1)} \left| \left< U \left(  \theta_{\mathbf{a}}^\mathbf{x} \right) \mid  (\mathbf{+}^{\otimes d} )  \right> \right|^2 \\
& = & 2^{-(d+1)} \left| \left< Z^K_{A_{d+1}}  \left( \theta_{\mathbf{a}}^\mathbf{x}  \right) \mid  (\mathbf{+}^{\otimes d} )  \right> \right|^2  \label{psiintermed}
\end{eqnarray}
where 
\begin{eqnarray}
    K & = & 1/4  + b_{d+1} +  \sum_{i=1}^d \left( \frac{y_i}{2} + b_i \right) x_i(-1)^{a_i} \\
    & = & 1/4 + b_{d+1} + \sum_{i=1}^d 
    \frac{ u_i v_i}{2}.
\end{eqnarray}
We have
\begin{eqnarray}
    Z^{K}_{A_{d+1}} \theta_{\mathbf{x}}^\mathbf{a} & = & \frac{1}{\sqrt{2}} \left( \left| a_1, a_2 ,  \ldots, a_d, 0 \right> + (-1)^{a_{d+1} + K} \left| a_1 \oplus x_1, a_2 \oplus x_2, \ldots, a_d \oplus x_d , 1 \right> \right), 
    \end{eqnarray}
    and
    \begin{eqnarray}
    a_{d+1} + K & = &  1/4 +  (a_{d+1} + b_{d+1}) + \sum_{i=1}^d 
    \frac{u_i v_i}{2}. 
\end{eqnarray}
By inspection, we find that $a_{d+1} + b_{d+1} \equiv (u_{d+1} v_{d+1})/2 - 1/2 \hskip0.1in (\textnormal{mod } 2)$, and therefore
\begin{eqnarray}
    Z^{K}_{A_{d+1}} \theta_{\mathbf{x}}^\mathbf{a} & = & \frac{1}{\sqrt{2}} \left( \left| a_1, a_2 ,  \ldots, a_d , 0 \right> + (-1)^{- 1/4 + \left< \mathbf{u} , \mathbf{v} \right>/2} \left| a_1 \oplus x_1, a_2 \oplus x_2,  \ldots, a_d \oplus x_d, 1 \right> \right)
\end{eqnarray}
Therefore, calculating explicitly the inner product in (\ref{psiintermed}),
\begin{eqnarray}
\psi^* \left(  M_\mathbf{x}^\mathbf{a} \otimes N_\mathbf{y}^\mathbf{b}  \right) \psi & = & 2^{-(d+1)} \left|  2^{-1/2} \cdot 2^{-(d+1)/2} + 2^{-1/2} (-1)^{- 1/4 + \left< \mathbf{u}, \mathbf{v} \right>/2} \cdot 2^{-(d+1)/2} \right|^2
\\
& = & 2^{-2(d+1)} \left|  \frac{ 1 + (-1)^{- 1/4 + \left< \mathbf{u}, \mathbf{v} \right>/2} }{\sqrt{2}} \right|^2 \\
& = & \left\{ \begin{array}{cl} 2^{-2(d+1)} \left( 1 + \frac{1}{\sqrt{2}} \right) & \textnormal{ if } \left< \mathbf{u}, \mathbf{v} \right> = 0 \textnormal{ or } 1 \\ \\
2^{-2(d+1)} \left( 1 - \frac{1}{\sqrt{2}} \right) & \textnormal{ if } \left< \mathbf{u}, \mathbf{v} \right> = 2 \textnormal{ or } 3
\end{array} \right.
\end{eqnarray}

For any fixed $\mathbf{x},\mathbf{y} \in \{ 0, 1 \}^{d+1}$ with $x_{d+1} = y_{d+1} = 1$, it easily checked that exactly half of the $2^{2d+2}$ pairs  $(\mathbf{a}, \mathbf{b})$ satisfy the condition $\left< \mathbf{u}, \mathbf{v} \right> \in \{ 0, 1 \}$.\footnote{The pairs that satisfy this condition are in one-to-one correspondence with the pairs that do not, via the NOT map on $a_{d+1}$.}  Therefore, if $\mathbf{x}, \mathbf{y}$ are given as input to Alice and Bob, then their output will achieve a score of $+1$ with probability $(1/2) (1 + 1/\sqrt{2})$ and a score of $-1$ with probability $(1/2) (1 - 1/\sqrt{2})$, yielding an expected score of
\begin{eqnarray}
    (1) (1/2) \left( 1 + \frac{1}{\sqrt{2}} \right) + 
    (-1) (1/2) \left( 1 - \frac{1}{\sqrt{2}} \right)
    & = & \frac{1}{\sqrt{2}},
\end{eqnarray}
as desired.
\end{proof}

Informally, we note that the strategy described in Proposition~\ref{prop:jent} also achieves an expected score of $\sqrt{2}/2$ at Game $\mathbf{J}'_{n}$.  However, we will not need to formalize that fact here.

\subsection{The Classical Value of $\mathbf{J}_d$}

\label{subsec:classval}

\begin{theorem}
    \label{thm:jn}
    The games $\{ \mathbf{J}_d \}_{d \geq 1 }$ satisfy
    \begin{eqnarray}
        \beta^c ( \mathbf{J}_d ) & \leq & \exp ( - \Omega ( d )). \label{classicalj}
    \end{eqnarray}
\end{theorem}

\begin{proof}
    Let $GHZ_4$ denote the $4$-player GHZ game (Definition~\ref{def:ghz}). 
    Our approach will be to prove the following inequality:
    \begin{eqnarray}
    \beta_c ( \mathbf{J}_d ) & \leq & 2 \omega_c ( GHZ_4^d )^{1/4},
    \end{eqnarray}
    from which inequality (\ref{classicalj}) then follows by Corollary~\ref{cor:fourghz}.

    Fix a positive integer $d$.  Let $(S, T)$ be a deterministic strategy for $\mathbf{J}_d$ such that the absolute value of the expected score achieved by $(S, T)$ is
    $\beta_c ( \mathbf{J}_d )$.  We first compute a simple expression for the expected score achieved by $(S, T)$.  
    Let $U \subseteq \mathbb{Z}_4^{d+1}$ be the set of all vectors of the form
    \begin{eqnarray}
      ( x_1 (-1)^{S_1 ( \mathbf{x })}, 
        x_2 (-1)^{S_2 ( \mathbf{x })}, \ldots, x_{d+1} (-1)^{S_{d+1} ( \mathbf{x })}  )
    \end{eqnarray}
    with $\mathbf{x} \in  \{ 0, 1 \}^d \times \{ 1 \}$.  Expressed differently, $U$ is the set of all possible values for the vector $\mathbf{u}$ in Figure~\ref{fig:gamej} (considered as an element of $\mathbb{Z}_4^{d+1}$) when the Game $\mathbf{J}_d$ is played with strategy $(S,T)$.  Similarly, let $V \subseteq \mathbb{Z}_4^{d+1}$ be the set of all vectors of the form
    \begin{eqnarray}
      ( y_1 + 2 T_1 ( \mathbf{y} ), y_2 + 2 T_2 ( \mathbf{y} ) , \ldots , y_{d+1} + 2 T_{d+1} ( \mathbf{y}) ).
    \end{eqnarray}
    with $\mathbf{y} \in \{ 0, 1 \}^{d} \times \{ 1 \}$.  The set $V$ is the set of residues mod $4$ of all possible values for the vector $\mathbf{v}$ in Figure~\ref{fig:gamej}.  (We note that, although $U$ is not a parity-balanced subset of $\mathbb{Z}_4^{d+1}$ according to Definition~\ref{def:parb}, it becomes a parity-balanced subset of $\mathbb{Z}_4^d$ if we drop the last coordinate of each vector that it contains. 
 The same is true of $V$.)  The expected score for the strategy $(S, T)$ can then be succinctly represented as
    \begin{eqnarray}
        \omega^c ( \mathbf{J}_d, (S, T) ) & = & \mathbf{E} \left[ \Real \left[ (1 - i ) i^{\mathbf{u} \cdot \mathbf{v} } \right] \mid \mathbf{u} \leftarrow U, \mathbf{v} \leftarrow V  \right].
        \label{omegac}
    \end{eqnarray}

    From equation~(\ref{omegac}), we can derive a formula  in terms of the Fourier transform.  Let $g,f \colon \mathbb{Z}_4^{d+1} \to \mathbb{C}$ be defined by
    \begin{eqnarray}
        g ( \mathbf{z} ) & = & 2^{-d/2} \delta_{\mathbf{z} \in U} \\
        f ( \mathbf{z} ) & = & 2^{-d/2} \delta_{\mathbf{z} \in V}, 
    \end{eqnarray}
    Note that $\left\| f \right\|_2 = \left\| g \right\|_2 = 1$.  We have
    \begin{eqnarray}
        \beta^c ( \mathbf{J}_d ) & = & \left| \omega^c ( \mathbf{J}_d, (S, T) ) \right|  \\
        & = & \left|  2^{-2d}  \Real \left[ (1-i)\sum_{\substack{\mathbf{u} \in U \\ \mathbf{v} \in V }} i^{\mathbf{u} \cdot \mathbf{v}} \right]  \right| \\
        & = & \left|  2^{-2d}  \Real \left[ (1-i) \sum_{\mathbf{u}, \mathbf{v} \in \mathbb{Z}_4^{n+1} } i^{\mathbf{u} \cdot \mathbf{v}} \delta_{\mathbf{v} \in V}  \delta_{\mathbf{u} \in U} \right] \right| \\
        & = &\left|  2^{-d}  \Real \left[ (1-i) \sum_{\mathbf{u}, \mathbf{v} \in \mathbb{Z}_4^{n+1} } i^{\mathbf{u} \cdot \mathbf{v}} f( \mathbf{u} ) g ( \mathbf{v})\right] \right|  \\
        & = & \left| 2^{-d}  \Real \left[ (1 - i )  \left< 2^{d+1} \hat{f} , g \right> \right] \right| \\
        & = & \left| 2  \cdot \Real \left[ (1 - i ) \left<\hat{f} , g \right> \right] \right|
    \end{eqnarray}
    which yields the inequality
    \begin{eqnarray}
        \beta^c ( \mathbf{J}_n ) 
        & \leq & 2 \sqrt{2}\left| \left<\hat{f} , g \right> \right|.
    \end{eqnarray}

    Next we apply Theorem~\ref{thm:uncertainty}.  We have
    \begin{eqnarray}
         \beta^c ( \mathbf{J}_d )  & \leq & 
        2 \sqrt{2}\left| \left<\hat{f} , g \right> \right| \\
        & \leq & 2 \sqrt{2} \left( \frac{ \left| \Supp f \right| \left| \Supp g \right| \nu ( f ) \eta (f ) }{\left| \mathbb{Z}_4^{d+1} \right|} \right)^{1/4}
    \end{eqnarray}
    Noting that $\left| \Supp f \right| = \left| \Supp g \right| = 2^d$, $\left| \mathbb{Z}_4^{d+1} \right| = 4^{d+1}$, and $\nu ( f ) = 1$ (Proposition~\ref{prop:unif}), we have
    \begin{eqnarray}
        \beta^c ( \mathbf{J}_d ) & \leq & 
        2 \sqrt{2} \left( \frac{\eta ( f )}{4} \right)^{1/4} \\
    & = & 2 \eta ( f )^{1/4} \\
    & = & 2 \eta ( V )^{1/4}.
    \end{eqnarray}
    Let $V'$ denote the subset of $\mathbb{Z}_4^d$ that arises from dropping the last coordinate of each vector in $V$.  
It is easy to see that $\eta ( V) \leq \eta ( V')$ (see Appendix~\ref{app:comment}).
The set $V'$ is a parity-balanced subset of $\mathbb{Z}_4^d$, and so by applying  Proposition~\ref{prop:parity}, we have
\begin{eqnarray}
         \beta^c ( \mathbf{J}_d ) & \leq & 
 2 \eta ( V )^{1/4} \\
    & \leq & 2 \eta ( V' )^{1/4} \\
    & \leq & 2 (\omega^c ( GHZ_4^d ))^{1/4},  \end{eqnarray}
as desired.
\end{proof}

We prove a stronger result for Game $\mathbf{J}'_d$.

\begin{theorem}
    \label{thm:jnprime}
    The games $\{ \mathbf{J}'_d \}_{d \geq 1 }$ satisfy
    \begin{eqnarray}
        \beta^c ( \mathbf{J}'_d ) & \leq & 2 \cdot (3/4)^{d/4}. 
    \end{eqnarray}
\end{theorem}

\begin{proof}
    The same proof as for Theorem~\ref{thm:jn} applies, with $\mathbf{J}_d$ replaced by $\mathbf{J}'_d$, and we merely need to observe that since Bob behaves sequentially in Game $\mathbf{J}'_d$, the set $V'$ is time-ordered  (Definition~\ref{def:timeparb}).  Therefore, applying Proposition~\ref{prop:timeparity},
    \begin{eqnarray}
          \beta^c ( \mathbf{J}'_d ) & \leq & 2 \eta ( V' )^{1/4} \\
& \leq & 2 \cdot (3/4)^{d/4},  \end{eqnarray}
    as desired.
\end{proof}

\section{Proofs of Quantumness}

\label{sec:pfs}

The setup in this section is based on \cite{alnawakhtha2024lattice}, and we use much of the same notation.  Throughout this section, $\lambda \in \mathbb{N}$ denotes a security parameter.  When we refer to an algorithm as ``polynomial-time,'' we mean polynomial in $\lambda$.  

Figure~\ref{fig:params} states the assumed constraints for seven parameters $(n, m, q, Q, \sigma, \tau, d)$, each of which is  a function of $\lambda$.   These constraints imply, in particular, that $q \geq \Omega ( n^2 \sigma)$.  An example of parameters that satisfy all of the constraints given in Figure~\ref{fig:params} is the following:
\begin{eqnarray*}
    n(\lambda) & = & \lambda \\
    q(\lambda) & = & \textnormal{an odd prime between } \lambda^3 \textnormal{ and } 2 \lambda^3 \\
    d(\lambda) & = & \lfloor \log \lambda \rfloor \\
    \sigma(\lambda) & = & \sqrt{\lambda},
\end{eqnarray*}
with $Q, m, \tau$ are set according to the formulas in Figure~\ref{fig:params}.

\begin{figure}
\fbox{\parbox{6in}{
\textit{Parameters:} \\ \\
\begin{tabular}{rl}
$q$: & modulus  \\
$Q$: & the binary length of integers mod $q$  \\
$m,n$: & matrix dimensions \\
$\sigma$: & standard deviation for Gaussian noise \\
$\tau$: & truncation parameter for Gaussian noise \\
$d$: & binary secret length
\end{tabular} \\ \\
\textit{Assumptions:}
\begin{itemize}
    \item $n = \lambda$
    \item $q$ is always an odd prime, and $q \leq \exp ( O ( \lambda ) )$
    \item $d \leq O (\log \lambda)$ and $d \leq \lambda$
    \item $Q = \lceil \log q \rceil$
    \item $m = (2Q+1)n$
    \item $\tau = \lfloor q/(4mQ) \rfloor$
    \item $\omega ( 1 ) \leq \sigma \leq o ( \tau / m)$
\end{itemize}
}
}
    \caption{Parameters for Section~\ref{sec:pfs}. $q, Q, m, n, d, \tau$ are positive-integer valued functions of $\lambda$, and $\sigma$ is a  function of $\lambda$ that takes on positive real values. }
    \label{fig:params}
\end{figure}

\begin{definition}
For any positive real number $s$, the \textbf{discrete Gaussian distribution} on $\mathbb{Z}$ with standard deviation $s$, denoted by $G( s )$, is the probability distribution on $\mathbb{Z}$ given by
\begin{eqnarray}
    G(s)(j) & = & \frac{ e^{-j^2/2 s^2}}{\sum_{j \in \mathbb{Z}} e^{-j^2/2 s^2}}.
\end{eqnarray}
If $t$ is a positive real number, the \textbf{truncated discrete Gaussian} $G ( s, t )$
is the distribution on $\mathbb{Z}$ that arises from restricting $G(s)$ to the subset $\left\{ j \mid \left| j \right| \leq t \right\}$ and normalizing.
\end{definition}

\subsection{Learning With Errors}

The Learning With Errors problem (LWE) has multiple variants.  We state a ``decisional'' version of the problem.   In the following, $\chi$ represents a probability distribution on $\mathbb{Z}$.  \\ 

\textbf{The $\LWE (n,q,\chi)$ Problem.}  Fix $s \leftarrow \mathbb{Z}_q^n$ and $b \leftarrow \{ 0, 1 \}$.  Let $\mathcal{D}_0$ be an oracle that outputs samples in $\mathbb{Z}_q^{n+1}$ of the form
\begin{eqnarray}
    (a, a \cdot s + e )
\end{eqnarray}
where $a \leftarrow \mathbb{Z}_q^n$ and $e \leftarrow \chi$, and let $\mathcal{D}_1$ be an oracle that outputs uniformly random samples in $\mathbb{Z}_q^{n+1}$.  Given oracle access to $\mathcal{D}_b$, compute $b$. \\ 

When we say that we assume that the $\LWE (n,q,\chi)$ problem is hard, we mean that we assume that any non-uniform polynomial-time randomized classical algorithm solves the $\LWE (n,q,\chi)$ problem with probability at most $\frac{1}{2} + \negl ( \lambda )$.

We will make use of random matrices in $\mathbb{Z}_q^{m \times n}$ that are generated with a trapdoor that allows for the efficient inversion of LWE samples.  The following theorem comes from \cite{micciancio2012trapdoors}, although we will refer instead to a version from \cite{alnawakhtha2024lattice} because it includes an explicit error term.

\begin{theorem}
    \label{thm:trap}
    There is a probabilistic polynomial-time algorithm $\GenTrap()$ and a deterministic polynomial-time algorithm $\Invert(A, t, v)$ satisfying the following conditions.
    \begin{itemize}
        \item The algorithm $\GenTrap()$ returns a pair $(A,t)$ where $A \in \mathbb{Z}_q^{m \times n}$.  The probability distribution of $A$ on $\mathbb{Z}_q^{m \times n}$ is within statistical distance $nQ2^{-n/2}$ from a uniform distribution.

        \item If $(A, t ) \leftarrow \GenTrap$ and $s \in \mathbb{Z}_q^n, e \in\mathbb{Z}_q^m$ are vectors such that $\left\| e \right\|_\infty \leq 2 \tau$, then $\Invert (A, t, As + e ) = s$.
    \end{itemize}
\end{theorem}

\begin{proof}
    See Subsection~3.5 and Appendix~C in \cite{alnawakhtha2024lattice}.
\end{proof}

\begin{remark}
    Theorem~\ref{thm:trap} implies that if $(A, t) \leftarrow \GenTrap ()$ and $v \in \mathbb{Z}_q^m$, there is at most one vector $s \in \mathbb{Z}_q^n$ such that $\left\| v - As \right\|_\infty \leq 2 \tau$.  If no such vector $s$ exists, then we will assume that
    $\Invert ( A, t, v ) = \bot$.
\end{remark}

We use Theorem~\ref{thm:trap} to build an encryption algorithm $\Encrypt$, shown in Figure~\ref{fig:encrypt}.
The following proposition is proved in Appendix~\ref{app:encrypt} by standard techniques.

\begin{proposition}
    \label{prop:encrypt}
    Assume that $LWE (n, q , G ( \sigma ))$ is hard.  Let $(x_1, \ldots, x_d ) \leftarrow \{ 0, 1 \}^d$ and
    \begin{eqnarray}
        (A, v, t ) \leftarrow \Encrypt (x_1 x_2 \ldots x_d ).
    \end{eqnarray}
    Then, the probability distribution of $(A,v, x_1 x_2 \ldots x_d )$  is computationally indistinguishable from the uniform distribution on $\mathbb{Z}_q^n \times \mathbb{Z}_q \times \{ 0, 1 \}^d$. $\Box$
\end{proposition}

\begin{figure}
\fbox{\parbox{6in}{
\textbf{Algorithm $\Encrypt( h)$:} \\ 
\\
\textit{Input:} A  bit string $h \in \{ 0, 1 \}^d$

\textit{Output:} A matrix $A \in \mathbb{Z}_q^{m \times n}$, a vector $v \in \mathbb{Z}_q^m$, and a classical register $t$.

\begin{enumerate}
    \item Compute $(A, t) \leftarrow \GenTrap()$.

    \item Compute $s \in \mathbb{Z}_q^n$ and $e \in \mathbb{Z}_q^m$ by sampling every entry of $s$ and $e$ independently from $G ( \sigma, \tau )$.

    \item Let $M \in \mathbb{Z}_q^n$ be a vector whose first $n-d$ entries are all zero and whose $n-d+j$th entry is $h_j$, and let $\gamma = (2s + M)$.  Compute
    \begin{eqnarray*}
        v: = A \gamma  + e \in \mathbb{Z}_q^m.
    \end{eqnarray*}
    Return $(A,v,t)$.
\end{enumerate}
}}
\caption{An encryption algorithm, in which $(A,v)$ is the ciphertext and $t$ is the secret key.}
\label{fig:encrypt}
\end{figure}

\subsection{The Central Games}

Game~$\mathbf{R}$, shown in Figure~\ref{fig:poq}, is a proof of quantumness based on an interaction between a prover (Alice) and a verifier (Referee).  Game~$\mathbf{R}$ is based on Game $\mathbf{J}_d$ (Figure~\ref{fig:gamej}), and Game~$\mathbf{R}'$ (Figure~\ref{fig:poqprime}) is a modification of Game~$\mathbf{R}$ that is based on Game~$\mathbf{J}'_d$ (Figure~\ref{fig:gamejprime}).

\begin{figure}
\fbox{\parbox{6in}{
\textbf{Game $\mathbf{R}$:} \\ 
\\
\textit{Parties:} Referee (verifier) and Alice (prover)

\begin{enumerate}
\item Referee samples random bit strings $\mathbf{x}, \mathbf{y} \in \{ 0, 1 \}^{d+1}$ subject to the constraint $x_{d+1} = y_{d+1} = 1$.  
    
    \item Referee computes  
    \begin{eqnarray*}
        (A, v, t) := \Encrypt (  x_1 x_2 \ldots x_d ).
    \end{eqnarray*}
    Referee sends $A$ and $v$ to Alice.

    \item Alice returns a vector $w \in \mathbb{Z}_q^m$ and an indexed set of bits $\{ \ell_j \}_{j \in S}$, where
    \begin{eqnarray*}
        S & = & \{ 1, 2, \ldots, nQ \} \smallsetminus \{ (n-d+1)Q, (n-d+2)Q, \ldots, nQ \}.
    \end{eqnarray*}

    \item Referee computes $z := \Invert ( A, t, w )$, $z' := \Invert ( A, t, w + v)$, and assigns
    \begin{eqnarray*}
        a_j & := & \left\{ \begin{array}{cl} 0 & \textnormal{ if } z_{n-d+j} \textnormal{ is even,} \\ \\
        1 & \textnormal{ otherwise.} \end{array} \right.
    \end{eqnarray*}
    for $i = 1, 2, \ldots, d$, and
    \begin{eqnarray*}
        a_{d+1} & := & \bigoplus_{j \in S} \left( \left( \left[z \right]_j \oplus \left[ z' \right]_j \right) \wedge \ell_j \right).
    \end{eqnarray*}
If either $\Invert$ procedure fails, Referee simply samples $\mathbf{a} \leftarrow \{ 0, 1 \}^{d+1}$.
    
\item Referee sends the vector $\mathbf{y}$ to Alice.  Alice returns a vector $\mathbf{b} \in \{ 0, 1\}^{n+1}$.

\item Referee computes $u_j := x_j(-1)^{a_j}$ and $v_j = y_j + 2 b_j$ for $ j = 1, 2, \ldots, n+1$.
If $\mathbf{u} \cdot \mathbf{v}$ is equal to $0$ or $1$ mod $4$, then Referee awards a score of $+1$.  Otherwise, Referee awards a score of $-1$.
\end{enumerate}
}}
\caption{An Interactive Proof of Quantumness}
\label{fig:poq}
\end{figure}

\begin{figure}
\fbox{\parbox{6in}{
\textbf{Game $\mathbf{R}'$:} \\ 
\\
Game $\mathbf{R}'$ is the same as Game $\mathbf{R}$ except that Step~5 is replaced with the following sequential interaction between Referee and Alice.
\begin{enumerate}
    \item[5.]
 For $j = 1, 2, 3, \ldots, d+1$, Referee sends $y_j$ to Alice and receives output $b_j \in \{ 0, 1\}^{d+1}$. \end{enumerate}
}}
\caption{A Modified Interactive Proof of Quantumness}
\label{fig:poqprime}
\end{figure}

\begin{figure}
\fbox{\parbox{6in}{
\textit{Step 3.} Alice prepares the state
\begin{eqnarray*}
 \phi & = & \frac{1}{\sqrt{2 q^n (2 \tau + 1 )^m}} \sum_{r \in \mathbb{Z}_q^n} \sum_{c \in \{ 0, 1 \}}
 \sum_{\substack{z \in \mathbb{Z}_q^m \\ \left\| z \right\|_\infty \leq \tau}}
  \left| r \right> \left| c \right> \left| Ar - cv + z \right>.
\end{eqnarray*}
Alice measures the third register of this state to obtain a state of the form $\psi \otimes \left| w \right>$ where $w \in \mathbb{Z}_q^m$ and $\psi$ is a pure state on $\mathbb{Z}_q^n \times \{ 0, 1 \}$.  She converts $\psi$ into a state $\psi'$ of $(nQ+1)$-qubits by applying binary representation to $\mathbb{Z}_q^n$ (see Remark~\ref{rem:bin}).  For each $j \in S$, she measures the $j$th qubit of $\psi'$ in the $X$-basis and records the result as $\ell_j \in \{ 0, 1 \}$.  She sends $w$ and $\{ \ell_j \}_{j \in S}$ to Referee. \\

\textit{Step 5.} For $j = 1, 2, \ldots, d$, Alice measures the $((n-d+j)Q)$th qubit of $\psi'$ in the $X$-basis if $y_j = 0$, or in $Y$-basis if $y_j = 1$, and records result as $a_j$.  Alice measures the $(nQ+1)$th qubit of $\psi'$ in the $(X+Y)/\sqrt{2}$ basis and records the result as $a_{d+1}$.  She sends $\mathbf{a}$ to Referee.
}}
\caption{A Quantum Strategy for Alice in Game $\mathbf{R}$.}
\label{fig:qstrat}
\end{figure}

We prove that a quantum prover can play Game $\mathbf{R}$ with a score approaching that of the optimal score for Game $\mathbf{J}_d$.

\begin{proposition}
    \label{prop:quantumgamer}
    Suppose that Alice behaves in Game $\mathbf{R}$ according to the strategy given in Figure~\ref{fig:qstrat}.  Then, her expected score is at least
    \begin{eqnarray}
        \frac{\sqrt{2}}{2} - o(1),
    \end{eqnarray}
    where $o(1)$ denotes a vanishing function of $\lambda$.
\end{proposition}

We will base our proof of Proposition~\ref{prop:quantumgamer} on the proof of Proposition~\ref{prop:jent}.  However, we must first address the possibility that the process followed by Alice in Step 3 fails to produce a proper claw-state.

\begin{proof}[Proof of Proposition~\ref{prop:quantumgamer}]
Let
\begin{eqnarray}
    z := \Invert ( A, t, w ) \hskip0.1in \textnormal{ and } \hskip0.1in z' := \Invert ( A, t, w + v ).
\end{eqnarray}
Either $\left\| A z - w \right\|_\infty \leq \tau$, or $\left\| A z' - w \right\|_\infty \leq \tau$, or both.  
The state $\psi$ is then (respectively) one of the following:
\begin{eqnarray}
    \left| z \right> \left| 0 \right> , \\
    \left| z' \right> \left| 1 \right> , \\
    \frac{1}{\sqrt{2}} \left( \left| z \right> \left| 0 \right> +
    \left| z' \right> \left| 1 \right> \right) .
\end{eqnarray}
Let $E$ be the event that both $\left\| A z - w \right\|_\infty \leq \tau$ and $\left\| A z' - w \right\|_\infty \leq \tau$.  Let $F$ be the event that for all $j \in \{ 1, 2, \ldots, n \}$, $\left| z_j \right| > \left| \gamma_j \right|$.\footnote{The absolute value here is taken within the ring $\mathbb{Z}_q$.  See Section~\ref{sec:prelim} for conventions regarding absolute value notation.}

If both $E$ and $F$ occur, then
\begin{eqnarray}
    \psi & = & 
\frac{1}{\sqrt{2}} \left( \left| z \right> \left| 0 \right> +
    \left| z' \right> \left| 1 \right> \right). 
\end{eqnarray}
and $z' = z + \gamma$, where $v = A \gamma + e$ and $\gamma,e$ are the vectors generated in the $\Encrypt$ algorithm (Figure~\ref{fig:encrypt}).  Since $\left| z_j \right| > \left| \gamma_j \right| \hskip0.05in \forall j$, the relationship $z' = z + \gamma$ holds not only over $\mathbb{Z}_q^n$, but also when $z', z, \gamma$ are considered as vectors in $\mathbb{Z}^n$.  Therefore, the following parity relationship holds:\footnote{See Remark~\ref{rem:bin} for an explanation of the base-$2$ notation used here.}
\begin{eqnarray}
    [z]_{Q, 2Q, \ldots, nQ} \oplus [z']_{Q, 2Q, \ldots, nQ} & = & [\gamma]_{Q, 2Q, \ldots, nQ} \\
    & = & 0^{n-d} || x_1 x_2 \ldots x_d.
\end{eqnarray}
The $(d+1)$-qubit state that remains for Alice at the end of Step $3$ is therefore precisely
\begin{eqnarray}
    \frac{1}{\sqrt{2}} \left( \left| a_1 a_2 \ldots a_d \hskip0.02in 0 \right> + 
    (-1)^{a_{n+1}} \left| (a_1 \oplus x_1) (a_2 \oplus x_2) \ldots (a_d \oplus x_d)  \hskip0.02in 1 \right> \right).
\end{eqnarray}
It follows by the same calculations as in the proof of Proposition~\ref{prop:jent} that, conditioned on $E \cap F$, Alice achieves an expected score of $\sqrt{2}/2$ at Game $\mathbf{R}$.  Therefore, to prove Proposition~\ref{prop:quantumgamer}, it will suffice to show that $\mathbf{P} ( E \cap F ) \geq 1 - o ( 1 )$.

Consider the sets
\begin{eqnarray}
\label{ima1}
    \Image ( A ) + [-\tau, \tau]^m
\end{eqnarray}
and
\begin{eqnarray}
\label{ima2}
    \Image ( A ) - v + [-\tau, \tau]^m    
\end{eqnarray}
For each vector $f \in \mathbb{Z}_q^n$, the set
\begin{eqnarray}
Af + [-\tau, \tau]^m    
\end{eqnarray}
overlaps with the set
\begin{eqnarray}
A(f+\gamma) - v + [-\tau, \tau]^m & = & Af + e + [-\tau, \tau]^m
\end{eqnarray}
and otherwise does not overlap with set~(\ref{ima2}).  The set $Af + [-\tau, \tau]^m$ is of size $(2 \tau + 1)^m$, while the overlap region is of size
\begin{eqnarray}
    \prod_{j=1}^m (2 \tau + 1 - |e_j| ).
\end{eqnarray}
Therefore,
\begin{eqnarray}
    \mathbf{P} [ E \mid e ] & = & \frac{\prod_{j=1}^m (2 \tau + 1 - |e_j| )}{(2 \tau + 1)^m} \\
    & = & \prod_{j=1}^m \left( 1 - \frac{  |e_j|}{2 \tau + 1 } \right) \\
    & \geq & 1 - \sum_j^m \frac{| e_j |}{2 \tau + 1} \\
    & = & 1 - \frac{ \left\| e \right\|_1 }{2 \tau + 1 }
\end{eqnarray}
So, using Lemma A.2 from \cite{alnawakhtha2024lattice},
    \begin{eqnarray}
    \mathbf{P} [ E ] & \geq & 1 - \mathbf{E} \left[ \frac{ \left\| e \right\|_1 }{2 \tau + 1 } \mid e \leftarrow G ( \sigma, \tau ) \right] \\
    & \geq & 1 - \frac{m \sigma}{2 \tau + 1},
    \end{eqnarray}
which, by the assumptions made about $\sigma$ in Figure~\ref{fig:params}, implies $\mathbf{P} [ E ] \geq 1 - o ( 1 )$.

We apply similar reasoning to derive a lower bound on $\mathbf{P}[F]$.  The vector $z$ is uniformly distributed over $\mathbb{Z}_q^n$, which is of size $q^n$, while the set
\begin{eqnarray}
    \left\{ z  \in \mathbb{Z}_q^n \mid \forall j \hskip0.05in \left| z_j \right| > \left| \gamma_j \right| \right\}
\end{eqnarray}
is of size 
\begin{eqnarray}
    \prod_{j=1}^m \left( q - 2 \left| \gamma_j \right| - 1 \right)
\end{eqnarray}
Therefore,
\begin{eqnarray}
    \mathbf{P} [ F \mid \gamma] & = & q^{-m} \cdot \prod_{j=1}^m \left( q - 2 \left| \gamma_j \right| - 1 \right)  \\
    & = & \prod_{j=1}^m \left( 1 - \frac{2 \left| \gamma_j \right| + 1}{q} \right) \\
    & \geq & 1 - \sum_{j=1}^m \frac{2 \left| \gamma_j \right| + 1}{q} \\
    & = & 1 - \frac{2 \left\| \gamma \right\|_1 + m}{q}.
\end{eqnarray}
From the $\Encrypt$ procedure, we have $\gamma = 2s + M$, where $\left\| M \right\|_1 \leq d$ and (using Lemma~A.2 from \cite{alnawakhtha2024lattice}) $\mathbf{E} [ \left\| s \right\|_1 ] \leq \sigma$.  Therefore,
\begin{eqnarray}
    \mathbf{P} [ F ] 
    & \geq  & 1 - \frac{2 \mathbf{E} \left[ \left\| \gamma \right\|_1 \right] + m}{q} \\
   & \geq  & 1 - \frac{4 \mathbf{E} \left[ \left\| s \right\|_1 \right] + 2 d + m}{q} \\
   & \geq & 1 - \frac{ 4 \sigma + 2 d + m }{q} \label{finalbound}
\end{eqnarray}
and the quantity (\ref{finalbound}) is easily to see to tend to $1$ as a function of $\lambda$ by the inequalities assumed in Figure~\ref{fig:params}.  
Therefore $\mathbf{P} ( E \cap F ) \geq 1 - \mathbf{P} ( \neg E ) - \mathbf{P} ( \neg F ) \geq 1 - o (1 )$.  This completes the proof.
\end{proof}

Lastly, we note that the strategy in Figure~\ref{fig:qstrat} is also a valid strategy for Game $\mathbf{R}'$, and so by the same proof we have the following.
\begin{proposition}
    \label{prop:quantumgamerprime}
    Suppose that Alice behaves in Game $\mathbf{R}'$ according to the strategy given in Figure~\ref{fig:qstrat}.  Then, her expected score is at least
    \begin{eqnarray}
        \frac{\sqrt{2}}{2} - o(1),
    \end{eqnarray}
    where $o(1)$ denotes a vanishing function of $\lambda$.
\end{proposition}

\subsection{Classical Soundness}

\label{subsec:classsound}

In this section, we use the expression $\Score ( \mathbf{x}, \mathbf{y}, \mathbf{a}, \mathbf{b} )$ to denote the score assigned to $( \mathbf{x}, \mathbf{y}, \mathbf{a}, \mathbf{b} )$ in Game~$\mathbf{J}_n$ (Figure~\ref{fig:gamej}).

\begin{theorem}
\label{thm:classicalgamer}
Suppose that Alice behaves in Game~$\mathbf{R}$ according to the model shown in Figure~\ref{fig:cstrat}.  Then, the bias of her strategy is upper bounded by $\exp ( - \Omega ( d )) + \negl (\lambda)$.
\end{theorem}

\begin{proof}
    Our proof method follows \cite{kalai2023quantum}.
    Let $s$ denote Alice's expected score in Game~$\mathbf{R}$. Consider Experiment $S_1$ shown in Figure~\ref{fig:exp1}, in which the process of generating Alice's responses is shared with a second player, Bob. 
    In Experiment $S_1$, Alice and Bob play $\mathbf{J}'_d$, but with a possible advantage: the referee has shared an encryption of Alice's input bits $x_1 x_2 \ldots x_d$ with both players in advance, and has also shared the trapdoor $t$ for the encryption matrix with Alice.
    The probability distribution of $(\mathbf{x}, \mathbf{y}, \mathbf{a}, \mathbf{b})$ is exactly that generated by Alice's original strategy in Game $\mathbf{R}$, and so
    \begin{eqnarray}
        s & = & \mathbf{E}_{S_1} \left[ \Score ( \mathbf{x}, \mathbf{y}, \mathbf{a}, \mathbf{b})  \right].
    \end{eqnarray}

    In Experiment $S_2$ (Figure~\ref{fig:exp2}), Alice skips computing her output string $\mathbf{a}$ directly, and instead predicts exactly what behavior Bob will exhibit on all of his possible inputs $\mathbf{y}$ (by running $\SecondResponse$ $2^d$ times).  Alice then chooses her output $\mathbf{b}$ so as to optimize the expected score against all possible inputs to Bob.  This new strategy can only increase the score achieved by Alice and Bob (since we have left Bob's behavior fixed and optimized Alice's behavior), and thus we have 
    \begin{eqnarray}
        s & \leq & \mathbf{E}_{S_2} \left[ \Score ( \mathbf{x}, \mathbf{y}, \mathbf{a}, \mathbf{b})  \right].
    \end{eqnarray}
    Importantly, since we have assumed $d \leq O ( \log \lambda )$, the process of Alice executing $\SecondResponse$ $2^d$ times only takes a polynomial amount of time.
    
    Finally, in Experiment $S_3$, there is one additional change: the Referee no longer provides Alice and Bob with an encryption $(A, v)$ of Alice's input bits $x_1 x_2 ... x_n$.  Instead, Referee merely shares a uniformly random $(A,v)$ with Alice and Bob.  By Proposition~\ref{prop:encrypt}, this change cannot be noticeable to Alice and Bob (who only use polynomial-time algorithms) and therefore its effect on the expected score is negligible:\footnote{The information transmitted from Referee to Alice and Bob collectively in Experiment $S_2$ consists of the string $\mathbf{x}$, its encryption $(A,v)$, and the string $\mathbf{y}$ (which is statistically independent of $\mathbf{x}$, $A$, and $v$).  If replacing the encryption $(A,v)$ with a uniformly random pair in $\mathbb{Z}_q^{m \times n} \times \mathbb{Z}_q^m$ caused a noticeable difference in the expected score of Experiment $S_2$, then Proposition~\ref{prop:encrypt} would be violated.}
    \begin{eqnarray}
        s & \leq & \mathbf{E}_{S_3} \left[ \Score ( \mathbf{x}, \mathbf{y}, \mathbf{a}, \mathbf{b} ) \right] + \negl ( \lambda ).
    \end{eqnarray}
    Experiment $S_3$ is merely Alice and Bob playing Game~$\mathbf{J}_d$ with a randomized classical strategy, and therefore by Theorem~\ref{thm:jn},
    \begin{eqnarray}
        s & \leq & \omega^c ( \mathbf{J}_d ) +  \negl ( \lambda ) \\
        & \leq & \exp ( - \Omega ( n ) ) + \negl ( \lambda ).
    \end{eqnarray}    
    An analogous argument shows that
    \begin{eqnarray}
        s & \geq & -\exp ( - \Omega ( n ) ) - \negl ( \lambda ),
    \end{eqnarray}
    which completes the proof.
\end{proof}

By repeating the same reasononing with Game~$\mathbf{R}$ replaced by Game~$\mathbf{R}'$ and Game~$\mathbf{J}_d$ replaced by Game~$\mathbf{J}'_d$, we obtain (using Theorem~\ref{thm:jnprime}):

\begin{theorem}
\label{thm:classicalgamerprime}
The bias of any classical polynomial-time strategy for Game~$\mathbf{R}'$ is upper bounded by
\begin{eqnarray}
    2 \cdot (3/4)^{-d/4} + \negl (\lambda).
\end{eqnarray}
\end{theorem}

\begin{figure}
\fbox{\parbox{6in}{
    \textit{Step 2:} Alice chooses the value of $coins$ uniformly at random and computes
    \begin{eqnarray*}
        (w, \{ \ell_j \},mem) \leftarrow \FirstResponse ( A, v, coins ),
    \end{eqnarray*}
    where $\FirstResponse$ is a non-uniform polynomial-time deterministic algorithm.  She sends $(w, \{ \ell_j \})$ to Referee. \\ 
    
    \textit{Step 5:} Alice computes
    \begin{eqnarray*}
        \mathbf{b} \leftarrow \SecondResponse ( \mathbf{y}, mem ),
    \end{eqnarray*}
    where $\SecondResponse$ is a non-uniform polynomial-time deterministic algorithm.  She sends $\mathbf{b}$ to Referee.
}}
    \caption{A model for the behavior of a classical prover (Alice) in Game~$\mathbf{R}$.  The register $coins$ denotes initial randomness and the register $mem$ denotes internal memory between responses.  Both registers are of polynomial size.}
\label{fig:cstrat}
\end{figure}

\begin{figure}
\fbox{\parbox{6in}{
\begin{enumerate}    
    \item Before the game begins, Alice samples $coins$ uniformly at random, and shares its value with Bob.
    
    \item Referee samples $\mathbf{x}$ and $\mathbf{y}$ uniformly at random from $\{ 0, 1 \}^d \times \{ 1 \}$.  
    
    \item Referee computes
    \begin{eqnarray*}
        (A,v,t) \leftarrow \Encrypt ( x_1 x_2 \dots x_d ).
    \end{eqnarray*}
    Referee sends $(A,v,t)$ to Alice and sends $(A,v)$ to Bob.
    
    \item Referee sends $\mathbf{x}$ to Alice.  Alice computes
    \begin{eqnarray*}
        (w, \{ \ell_j \}, mem ) \leftarrow \FirstResponse (A, v, coins ),
    \end{eqnarray*}
    and then computes the vector $\mathbf{a} \in \{ 0, 1 \}^{d+1}$ via the procedure in Step 4 of Figure~\ref{fig:poq}.  She sends $\mathbf{a}$ to Referee.

    \item Referee sends $\mathbf{y}$ to Bob.  Bob also computes     
    \begin{eqnarray*}
        (w, \{ \ell_j \}, mem ) \leftarrow \FirstResponse (A, v, coins ),
    \end{eqnarray*}
    and computes
    \begin{eqnarray*}
        \mathbf{b} \leftarrow \SecondResponse (\mathbf{y}, mem )
    \end{eqnarray*}
    and sends $\mathbf{b}$ to Referee.
\end{enumerate}
}}

    \caption{Experiment $S_1$, in which Alice and Bob play Game~$\mathbf{J}_d$ with extra advice from the referee.}
\label{fig:exp1}
\end{figure}

\begin{figure}
\fbox{\parbox{6in}{
\begin{enumerate}    
    \item Before the game begins, Alice samples $coins$ uniformly at random, and shares its value with Bob.
    
    \item Referee samples $\mathbf{x}$ and $\mathbf{y}$ uniformly at random from $\{ 0, 1 \}^d \times \{ 1 \}$.
    
    \item Referee computes
    \begin{eqnarray*}
        (A,v,t) \leftarrow \Encrypt ( x_1 x_2 \dots x_d ).
    \end{eqnarray*}
    Referee sends $(A,v)$ to both Alice and Bob.
    
    \item Referee sends $\mathbf{x}$ to Alice.   Alice computes 
    \begin{eqnarray*}
        (w, \{ \ell_j \}, mem ) \leftarrow \FirstResponse (A, v, coins ),
    \end{eqnarray*} and then computes a vector $\mathbf{a} \in \{ 0, 1\}^{d+1}$ which maximizes the expectation
    \begin{eqnarray*}
        \mathbf{E} \left[ \Score ( \mathbf{x}, \mathbf{y}, \mathbf{a}, \SecondResponse (\mathbf{y}, mem )) \mid \mathbf{y} \leftarrow \{ 0, 1 \}^d \times \{ 1 \} \right].
    \end{eqnarray*}
    She sends $\mathbf{a}$ to the Referee.

    \item Referee sends $\mathbf{y}$ to Bob.   Bob also computes     
    \begin{eqnarray*}
        (w, \{ \ell_j \}, mem ) \leftarrow \FirstResponse (A, v, coins ),
    \end{eqnarray*}
    and computes
    \begin{eqnarray*}
        \mathbf{b} \leftarrow \SecondResponse (\mathbf{y}, mem )
    \end{eqnarray*}
    and sends $\mathbf{b}$ to Referee.
\end{enumerate}
}}

    \caption{Experiment $S_2$.  Since we have assumed that $d$ is $O ( \log \lambda)$, all of the procedures in this experiment can be performed in polynomial time.}
\label{fig:exp2}
\end{figure}

\begin{figure}
\fbox{\parbox{6in}{
Experiment $S_3$ is the same as Experiment $S_2$, except that Step 3 is replaced with the following.

\begin{enumerate}    
    \item[3.] Referee samples $(A, v) \leftarrow \mathbb{Z}_q^{m \times n } \times \mathbb{Z}_q^m$ and sends $(A,v)$ to both Alice and Bob.
\end{enumerate}
}}

    \caption{Experiment $S_3$.  The advice provided to Alice and Bob by the Referee is now uniformly random.}
\label{fig:exp3}
\end{figure}

\section{Numerical Calculation}

\label{sec:numerical}

The proof of Theorem~\ref{thm:classicalgamer} showed that if there is a classical cheating strategy for Game~$\mathbf{R}$ or Game $\mathbf{R}'$ that has a sufficiently high expected score, then we can use the algorithms from the cheating strategy to perform a distinguishing attack on $\Encrypt$.  (Via the proof of Proposition~\ref{prop:encrypt}, this then implies a solver for the $\LWE_{n, q , G( \sigma) }$ problem.)  The question then becomes: how large must the expected score be in order to make this attack significant?  In other words, how large must the expected score be to yield an attack on LWE that is far more efficient than the best known classical attacks on LWE?
In this section, we sketch some additional theory for answering this question and we give a preliminary calculation.

We begin by optimizing some of the constructions from subsection~\ref{subsec:classsound}.  If  $\mathbf{x} \in \{ 0, 1 \}^n \times \{ 1 \}$ is an input string for the first player in Game $\mathbf{J}_d$, then $\BestScore$ (Figure~\ref{fig:bestscore}) is an algorithm to compute the best possible expected score that the first player can achieve by optimally choosing their output $\mathbf{a}$.  The algorithm BestScore also takes as input a list of second-player input-output pairs $(\mathbf{y}_j, \mathbf{b}_j)$ that $\mathbf{x},\mathbf{a}$ may be scored against.  These pairs can either be a complete list of the inputs and corresponding outputs for the second player, or a random sample thereof.  (The maximization procedure assumes that the inputs and outputs for the second player are uniformly sampled from these pairs.)  The algorithm $\BestScore$ uses as a subroutine the algorithm $\DecodeError$ (Figure~\ref{fig:decodeerror}), which is an algorithm that computes the Hamming distance between a vector $\mathbf{w}$ and the image of a given matrix $B$.

\begin{figure}
    \centering
    \fbox{\parbox{5.5in}{\textbf{Algorithm} $\BestScore (\mathbf{x}, ((\mathbf{y}_j, \mathbf{b}_j))_j)$: \\

\textit{Input:} A bit string $\mathbf{x} \in \{ 0, 1 \}^{d+1}$ and a sequence  of bit string pairs $(\mathbf{y}_1, \mathbf{b}_1 ),
\ldots, (\mathbf{y}_c, \mathbf{b}_c ) \in \{ 0, 1 \}^{d+1} \times \{ 0, 1 \}^{d+1}$.

\textit{Output:} A real number between $-1$ and $+1$.

\begin{enumerate}
\item Let $B \in \mathbb{Z}_2^{ c \times (d+1)}$ be the binary matrix whose $j$th row is $\mathbf{x} \wedge \mathbf{y}_j$. 

\item
Compute the vector $\mathbf{w} \in \mathbb{Z}_2^c$ defined by
\begin{eqnarray*}
    w_j & = & \left\{ \begin{array}{cl} 0 & \textnormal{ if } \left< \mathbf{x} ,
    \mathbf{y}_j + 2 \mathbf{b}_j \right> \equiv 0 \textnormal{ or } 1 \textnormal{ } (\textnormal{mod } 4 ) \\
    \\
    1 & \textnormal{ otherwise.} \end{array} \right.
\end{eqnarray*}

\item Return $1 - 2 \cdot \DecodeError( B, \mathbf{w} )/c$.
\end{enumerate}    
    }}
    \caption{The algorithm $\BestScore$.}
    \label{fig:bestscore}
\end{figure}

\begin{figure}
    \centering
    \fbox{\parbox{5.5in}{\textbf{Algorithm} $\DecodeError (B, \mathbf{w}))$: \\

\textit{Input:} A matrix $B \in \mathbb{Z}_2^{(d+1)\times c}$ and a vector $\mathbf{w} \in \mathbb{Z}_2^c$, where $c$ is a positive integer.

\textit{Output:} A nonnegative integer.

\begin{enumerate}
\item Let $\zeta = c$.

\item For $j = 0, 1, 2, \ldots , 2^{d+1} - 1$,

\begin{itemize}
    \item Let $\mathbf{z} \in \mathbb{Z}_2^{d+1}$ be the binary representation of $j$, and compute \[ \zeta \leftarrow \min \left\{ \zeta, \left\| B \mathbf{z} - \mathbf{w} \right\|_1 \right\}. \]
\end{itemize}

\item Return $\zeta$.
\end{enumerate}    
    }}
    \caption{The algorithm $\DecodeError$.}
    \label{fig:decodeerror}
\end{figure}

\begin{figure}
    \centering
    \fbox{\parbox{5.5in}{\textbf{Experiment} $E$: \\

\textit{Parties:} Alice, Referee.

\begin{enumerate}
\item Referee samples $b \leftarrow \{ 0, 1 \}$ and $\mathbf{x} \leftarrow \{ 0, 1 \}^d \times \{ 1 \}$.  Referee sends $\mathbf{x}$ to Alice.

\item If $b = 0$, then Referee computes 
\[ (A, v, t ) \leftarrow \Encrypt ( x_1 x_2 \ldots x_d ) \].  If $b = 1$, Referee samples $(A,v) \leftarrow \mathbb{Z}_q^{m  \times n} \times \mathbb{Z}_q^m$.  Referee sends $(A,v)$ to Alice.

\item Alice samples the value of $coins$ uniformly at random and computes
\begin{eqnarray*}
    ( w, \{ \ell_j \}, mem) \leftarrow \FirstResponse ( A, v, coins ).
\end{eqnarray*}

\item Alice puts the elements of $\{ 0, 1\}^d \times \{ 1 \}$ into a sequence $\mathbf{y}_1, \ldots, \mathbf{y}_{2^d}$.  Alice computes
\begin{eqnarray*}
    \mathbf{b}_j \leftarrow \SecondResponse ( \mathbf{y}_j , mem )
\end{eqnarray*}
for $j = 1, 2, \ldots, 2^d$.

\item Alice computes
\begin{eqnarray*}
    \rho \leftarrow \BestScore(\mathbf{x}, ((\mathbf{y}_j, \mathbf{b}_j))_j).
\end{eqnarray*}

\item Alice samples $r \in \{ -1, 1 \}$ according to a distribution that has expected value $\rho$.  

\item If $r = +1$, Alice returns the bit $0$, and if $r =  -1$, Alice returns $1$. 
\end{enumerate}

}}
\caption{Experiment $E$. Alice tries to guess the value of $b$.} 
\label{fig:expe}
\end{figure}

\begin{figure}
    \centering
    \fbox{\parbox{5.5in}{\textbf{Experiment} $E'$: \\

Experiment $E'$ is the same as Experiment $E$, except that there is a supplied parameter $\alpha \in \{ 1, 2, \ldots, 2^d \}$, and Step 4 proceeds as follows:

\begin{enumerate}
\item[4.] Alice samples $\mathbf{y}_1, \mathbf{y}_2, \ldots, \mathbf{y}_\alpha \leftarrow \{ 0, 1 \}^d \times \{ 1 \}$ (from a uniform distribution).  Alice computes
\begin{eqnarray*}
    \mathbf{b}_j \leftarrow \SecondResponse ( \mathbf{y}_j , mem )
\end{eqnarray*}
for $j = 1, 2, \ldots, \alpha$.
\end{enumerate}

}}
\caption{Experiment $E'$. A more efficient way to approximate the outcome of Experiment $E$.} 
\label{fig:expeprime}
\end{figure}

Suppose that $(\FirstResponse, \SecondResponse)$ is a classical strategy (see Figure~\ref{fig:cstrat}) which achieves an expected score of
$\omega^c ( \mathbf{J}_d ) + \epsilon$. 
In Experiment~$E$ (Figure~\ref{fig:expe}), a Referee samples a bit $b \leftarrow \{ 0, 1 \}$.  If $b = 0$, then Referee and Alice conduct a simulation of Experiment~$S_2$ (Figure~\ref{fig:exp2}), and if $b = 1$, then Referee and Alice conduct a simulation of Experiment~$S_3$.  The resulting score from the simulation is denoted by $r$.  By the reasoning from subsection~\ref{subsec:classsound}, we have
\begin{eqnarray}
    \mathbf{E} [ r \mid b = 0 ] & \geq & \omega^c ( \mathbf{J}_d ) + \epsilon, \\
    \mathbf{E} [ r \mid b = 1 ] & \leq & \omega^c ( \mathbf{J}_d ).
\end{eqnarray}

Alice's goal is to guess the value of $b$ (which is equivalent to a distinguishing attack on $\Encrypt$).
At the end of the protocol, if $r = +1$, she guesses that $b = 0$, and if $r = -1$, she guesses that $b = 1$.   The result is that she guesses $b$ correctly with probability $\frac{1}{2} + \epsilon/2$.  

The running time for Experiment $E$ is roughly
\begin{eqnarray}
    T_{E} & \approx & T_{\DecodeError, 2^d\times d} + T_\FirstResponse + 2^d T_\SecondResponse,
\end{eqnarray}
where $T_*$ denotes the running time for the subscripted procedure, and we have written $T_{\DecodeError, 2^d \times d}$ for the running time of $\DecodeError$ when it is run with $c = 2^d$.

However, we can do better.  Experiment~$E'$ (Figure~\ref{fig:expeprime}) is the same as Experiment~$E$ except that a positive integer parameter $\alpha$ is given, and  only a randomly chosen sequence of  input-output pairs $(\mathbf{y}_j, \mathbf{b}_j)$ are used to estimate the score and provide the probability distribution for $r \in [-1, 1 ]$.  By Proposition~\ref{prop:maxsamp}, this does not change the expected value of $r$ by more than
\begin{eqnarray}
    \frac{2 + \sqrt{ \log \alpha + 2 \log \left| \mathbb{Z}_2^d \right|}}{\sqrt{\alpha}}
     & = & \frac{2 + \sqrt{ \log \alpha + 2d }}{\sqrt{\alpha}}.
\end{eqnarray}
Therefore, if we choose $\alpha$ to be a large multiple of $d/\epsilon^2$, then a significant positive gap will remain between the expected value of $r$ and the expected value of $r'$ in Experiment~$E'$.  A running-time expression for $T_{E'}$ is then given by:
\begin{eqnarray}
    \label{eprimexp}
    T_{E'} & \approx & T_{\DecodeError, \alpha \times d} + T_\FirstResponse + \alpha \cdot T_\SecondResponse,
\end{eqnarray}
The same reasoning carries over when we replaced Game $\mathbf{J}_d$ with Game $\mathbf{J}'_d$ and Game $\mathbf{R}$ with Game $\mathbf{R}'$, and then one can take advantage of the  explicit upper bound for $\omega^c ( \mathbf{J}'_d )$ given by Theorem~\ref{thm:jnprime}.

Take, for example, the parameters
\begin{eqnarray}
    d & = & 40 \\
    \epsilon & = & 0.05 \\
    \alpha & = & 400,000.
\end{eqnarray}
Then,
\begin{eqnarray}
    \omega^c \left( \mathbf{J}'_d \right) & \leq & 2 \cdot (3/4)^{d/4} < 0.1127
\end{eqnarray}
Suppose that $(\FirstResponse, \SecondResponse)$ 
is a cheating strategy for Game $\mathbf{R}'$ that wins with probability $\omega^c \left( \mathbf{J}'_d \right) + \epsilon$.  Then, in Experiment $E'$,
\begin{eqnarray}
\mathbf{E} [ r \mid b = 0 ] &  \geq & 
\omega^c \left( \mathbf{J}'_d \right) + \epsilon 
-     \frac{2 + \sqrt{ \log \alpha + 2d }}{\sqrt{\alpha}} \\ & \approx & \omega^c \left( 
\mathbf{J}'_d \right) + 0.05 - \frac{2 + 
\sqrt{ 18.6096 + 80}}{632.45} \\
& \approx & \omega^c \left( 
\mathbf{J}'_d \right) + 0.05 - 0.01886 \\
& =  & \omega^c \left( 
\mathbf{J}'_d \right) + 0.03114
\end{eqnarray}
and
\begin{eqnarray}
\mathbf{E} [ r \mid b = 1 ] &  \leq & 
\omega^c \left( \mathbf{J}'_d \right)  
+     \frac{2 + \sqrt{ \log \alpha + 2d }}{\sqrt{\alpha}} \\ & \approx & \omega^c \left( 
\mathbf{J}'_d \right)  +  0.01886 .
\end{eqnarray}
Therefore, Experiment $E'$ breaks the security of $\Encrypt$.  The running time for Experiment $E'$ is roughly
\begin{eqnarray}
    T_{E'} & \approx & T_{\DecodeError,  400000 \times 40 } + T_\FirstResponse + 400000 \cdot T_\SecondResponse,
\end{eqnarray}
Moreover, the number of nonzero columns in the matrix $B$ that is passed to $\DecodeError$ is no more than the Hamming weight of $\mathbf{x}$, which, except with probability $\leq 0.0012$, is no more than $30$.  If we ignore the zero columns of $B$, the running time can be roughly bounded by
\begin{eqnarray}
T_{\DecodeError, 400000 \times 30 } + T_\FirstResponse + 400000 \cdot T_\SecondResponse.
\end{eqnarray}
Multiplying a $400,000 \times 30$ matrix by a length $30$ vector could take roughly $2 \cdot 400,000 \cdot 30 \approx 2^{24.6}$ bit operations, and this would be repeated $2^{30}$ times for a total of roughly $2^{55}$ bit operations.\footnote{For comparison, a computer running at $4$ GHz would complete roughly $2^{57}$ cycles in one year.  The value $400,000$ is comparable to the number of seconds in a five-day period.}  

Since \begin{eqnarray} \omega^c ( \mathbf{J}'_d ) + \epsilon &  \leq & 2 \cdot (3/4)^{d/4} + \epsilon \\
& < & 0.1127 + 0.05 \\
& = & 0.1617,
\end{eqnarray}
a score of at least $0.1617$ at Game $\mathbf{R}'$ is sufficient to imply such an attack on the $\Encrypt$ protocol.

\appendix

\section{Supporting Proofs}

\subsection{Proof of Corollary~\ref{cor:fourghz}}

\label{app:fourghz}

We need only to show that for any $d \in \mathbb{N}$,
\begin{eqnarray}
    \omega^c ( GHZ_4^d ) & \leq & \omega^c ( GHZ_3^d ).
\end{eqnarray}

Fix $d \in \mathbb{N}$, and let $(S, T, U, V)$ be a deterministic strategy for $GHZ_4^d$ which achieves the optimal expected score.  For any bit strings $\mathbf{t}, \mathbf{z} \in \{ 0, 1 \}^d$, 
\begin{eqnarray}
    F_\mathbf{t} ( \mathbf{z} ) 
    & = & U ( \mathbf{t} ) \oplus V ( \mathbf{t} \oplus \mathbf{z} ) \oplus (\neg \mathbf{z} \wedge \mathbf{t} ).
\end{eqnarray}
Suppose that Alice, Bob, and Charlie play $GHZ_3^d$ as follows:
\begin{enumerate}
    \item Charlie samples $\mathbf{t} \leftarrow \{ 0, 1 \}^d$. 

    \item Upon receiving their input strings, Alice, Bob, and Charlie use the functions $S$, $T$, and $F_{\mathbf{t}}$ respectively to compute their outputs.
\end{enumerate}
By direct computation, one can see that the expected score achieved by this strategy is the same as the expected score achieved by $(S, T, U, V)$ at $GHZ_4^d$, which is $\omega^c ( GHZ_4^d)$.  Therefore, at least one of the deterministic strategies $(S, T, F_\mathbf{t})$ must achieve an expected score at $GHZ_3^d$ of at least $\omega^c ( GHZ_4^d)$.  This completes the proof.

\subsection{The Classical Value of $GHZ_4$}

\label{app:ghzscore}

We follow \cite{mermin1990extreme}.
Suppose that $(F_1, F_2, F_3, F_4)$ is a deterministic strategy for $GHZ_4$, and let
\begin{eqnarray}
    v_j & = & \frac{(-1)^{F_j(0)} + i \cdot (-1)^{F_j(1)} }{\sqrt{2}} \in \mathbb{C}.
\end{eqnarray}
Then, the expected score achieved by this strategy is
\begin{eqnarray}
    \frac{1}{2} + \frac{\Real [ v_1 v_2 v_3 v_4 ]}{4}
\end{eqnarray}
Since each $v_i$ is a unit-length complex number, the quantity above obviously cannot exceed $3/4$.  And, setting $F_1 ( x ) = F_2(x ) = 0$ and $F_3 (x ) = F_4 (x ) = x$ achieves a score of $3/4$.

\subsection{Comment on the proof of Theorem~\ref{thm:jn}}

\label{app:comment}

We have
\begin{eqnarray}
    \eta ( V ) & = & \frac{ \mathbf{P} [\mathbf{v}_1 + \mathbf{v}_2 = \mathbf{v}_3 + \mathbf{v}_4 \mid \mathbf{v}_1, \mathbf{v}_2, \mathbf{v}_3, \mathbf{v}_4 \leftarrow V ]}{\mathbf{P} [\mathbf{v}_1 = \mathbf{v}_2 \mid \mathbf{v}_1, \mathbf{v}_2 \leftarrow V ] } \label{thefrac}
\end{eqnarray}
The set $V$ contains $2^n$ elements, one from each coset
\begin{eqnarray}
    (x_1, x_2, \ldots, x_n, 1 ) + 2 \mathbb{Z}_4^n
\end{eqnarray}
with $x_1, x_2, \ldots, x_n \in \{ 0, 1 \}$.
The set $V'$ is obtained by dropping the last coordinate of each vector in $V$.  
This operation has no effect on the denominator in (\ref{thefrac}) (which is $2^{-n}$ for both $V$ and $V'$) and cannot decrease the numerator in (\ref{thefrac}), and so we find that
    \begin{eqnarray}
        \eta ( V ) & \leq & \eta ( V').
    \end{eqnarray}

\subsection{Proof of Proposition~\ref{prop:encrypt}}

\label{app:encrypt}

We define an alternative algorithm $\FakeEncrypt ( h )$, which is the same as $\Encrypt ( h )$ except that the matrix $A$ is generated uniformly at random instead of from $\GenTrap()$, and that vectors $s,e$ are generated according to the distribution $G(\sigma)$ instead of the truncated distribution $G( \sigma, \tau)$.  (We call the procedure $\FakeEncrypt$ because there is no decryption key in the output.)

\begin{figure}
\fbox{\parbox{6in}{
\textbf{Algorithm $\FakeEncrypt( h)$:} \\ 
\\
\textit{Input:} A bit string $h \in \{ 0, 1 \}^d$

\textit{Output:} A matrix $A \in \mathbb{Z}_q^{m \times n}$ and a vector $v \in \mathbb{Z}_q^m$.

\begin{enumerate}
    \item Sample $A \leftarrow \mathbb{Z}_q^{m \times n}$.

    \item Compute $s \in \mathbb{Z}_q^n$ and $e \in \mathbb{Z}_q^m$ by sampling every entry of $s$ and $e$ independently from $G ( \sigma )$.

    \item Let $M \in \mathbb{Z}_q^n$ be a vector whose first $n-d$ entries are all zero and whose $n-d+j$th entry is $h_j$, and let $\gamma = (2s + M)$.  Compute
    \begin{eqnarray*}
        v: = A \gamma  +  e \in \mathbb{Z}_q^m.
    \end{eqnarray*}
    Return $(A,v)$.
\end{enumerate}
}}
\caption{A fake encryption algorithm with no decryption key.}
\label{fig:fakeencrypt}
\end{figure}

Since $\sigma \leq o ( \tau / m )$ (see Figure~\ref{fig:params}), $G ( \sigma )$ is statistically indistinguishable from $G ( \sigma, \tau )$.  And by Theorem~\ref{thm:trap}, the distribution of the matrix $A$ output by $\GenTrap()$ is statistically indistinguishable from a uniform distribution.  Therefore, it suffices to prove that if $x_1 x_2 \ldots x_d \leftarrow \{ 0, 1 \}^d$ and 
\begin{eqnarray}
    (A, v ) \leftarrow \FakeEncrypt (x_1 x_2 \ldots x_d),
\end{eqnarray}
then the distribution of $(A, v, x_1 x_2 \ldots x_d)$ is computationally indistinguishable from uniform.  

We consider the Normal-Form Learning With Errors Problem ($\NFLWE$), which is the same as the $\LWE$ problem except that the secret vector $s$ is chosen via the noise distribution $\chi$ instead of from a uniform distribution.

\textbf{The $\NFLWE (n,q,\chi)$ Problem.}  Fix $s \leftarrow \chi^n$ and $b \leftarrow \{ 0, 1 \}$.  Let $\mathcal{D}_0$ be an oracle that outputs samples in $\mathbb{Z}_q^{n+1}$ of the form
\begin{eqnarray}
    (a, a \cdot s + e )
\end{eqnarray}
where $a \leftarrow \mathbb{Z}_q^n$ and $e \leftarrow \chi$, and let $\mathcal{D}_1$ be an oracle that outputs uniformly random samples in $\mathbb{Z}_q^{n+1}$.  Given oracle access to $\mathcal{D}_b$, compute $b$. \\ 

The hardness of $\LWE(n,q,G ( \sigma))$ implies the hardness of $\NFLWE ( n,q,G( \sigma ))$ (see subsection 4.2.1 in \cite{peikert2016decade}).  Therefore a sample of the form
\begin{eqnarray}
    (B, B s + e)
\end{eqnarray}
where $B \leftarrow \mathbb{Z}_q^{m \times n}$, $s \leftarrow G(\sigma )^n$, $e \leftarrow G ( \sigma )^m$, is computationally indistiguishable from uniform.  Since multiplication by $2$ is bijective on $\mathbb{Z}_q$, we find that 
\begin{eqnarray}
    (2^{-1} B, B s + e)
\end{eqnarray}
is likewise indistinguishable from uniform.  Letting $A = 2^{-1} B$, we have that
\begin{eqnarray}
    (A, 2A s + e),
\end{eqnarray}
where $A \leftarrow \mathbb{Z}_q^n$ and $s,e$ are sampled as before, is indistinguishable from uniform.  Note that the vector $v$ produced by $\FakeEncrypt$ is created by adding the vector $z := 2As + e$ to the vector $AM$.  
The probability distribution of
\begin{eqnarray}
    (A, v, x_1 x_2 \ldots x_d )
\end{eqnarray}
where $x_1 x_2 \ldots x_d \leftarrow \{ 0, 1 \}^d$ and $(A, v ) \leftarrow \Encrypt ( x_1, x_2, \ldots, x_d )$ is therefore computationally indistiguishable from a distribution
of the form \begin{eqnarray}
    (A, z + AM, x_1 x_2 \ldots x_d ),
\end{eqnarray}
where $x_1 x_2 \ldots x_d \leftarrow \{ 0, 1 \}^d$, $M = 0^{n-d} || \mathbf{x}$, $A \leftarrow \mathbb{Z}_q^{m \times n}$, $z \leftarrow \mathbb{Z}_q^m$.  Since the latter distribution is clearly uniform, this completes the proof.

\subsection{A Proposition on Sampling}

\begin{proposition}
    \label{prop:maxsamp}
    Let $S, T$ be be finite sets, and let $F \colon S \times T \to [-1, 1]$ be a function.  Let
    \begin{eqnarray}
        \Lambda & := & \max_{s \in S} \frac{ \sum_{t \in T} F ( s,t)}{\left| T \right|}
    \end{eqnarray}
    Let $\alpha$ be a positive integer.  Suppose that independent samples $t_1, \ldots, t_\alpha \leftarrow T$ are chosen, and
    \begin{eqnarray}
        \Lambda' & := & \max_{s \in S} \frac{  F ( s,t_1) + F ( s, t_2) + \cdots + F ( s, t_\alpha )}{\left| T \right|}
    \end{eqnarray}
    Then,
    \begin{eqnarray}
        \left| \mathbf{E} [ \Lambda' ] - \Lambda \right| & \leq & \frac{2 + \sqrt{ \log \alpha + 2 \log \left| S \right|}}{\sqrt{\alpha}}.
    \end{eqnarray}
\end{proposition}

\begin{proof}
    Let $\zeta > 0$.  For any $s \in S$, let 
    \begin{eqnarray}
        \Lambda_s & := & \frac{ \sum_{t \in T} F ( s,t)}{\left| T \right|}
    \end{eqnarray}
    and 
   \begin{eqnarray}
        \Lambda'_s & := & \frac{  F ( s,t_1) + F ( s, t_2) + \cdots + F ( s, t_\alpha )}{\left| T \right|}
    \end{eqnarray}
    By Hoeffding's inequality, for any fixed $s$, the probability of the event $\left| \Lambda_s - \Lambda'_s \right| \geq \zeta$ is no more than
    \begin{eqnarray}
        \exp ( - \alpha \zeta^2 / 2).
    \end{eqnarray}
    Therefore, the probability of the event $\exists_{s \in S} \left| \Lambda'_s - \Lambda_s \right| \geq \zeta$ is no more than 
    \begin{eqnarray}
        \label{expquantity}
        \left| S \right| \exp ( - \alpha \zeta^2 / 2).
    \end{eqnarray}
    The probability that $\left| \Lambda' - \Lambda \right| \geq \zeta$ is likewise no greater than (\ref{expquantity}).  
    In the event that the quantity $\left| \Lambda' - \Lambda \right|$ is greater than $\zeta$, it still cannot be more than $2$. Therefore we have
    \begin{eqnarray}
        \left| \mathbf{E} [ \Lambda' ] - \Lambda \right| & \leq & \zeta + 2 \left| S \right| \exp ( - \alpha \zeta^2 / 2)
    \end{eqnarray}
    Let
    \begin{eqnarray}
        \zeta & = & \sqrt{ \frac{\log \left( \alpha  \left| S \right|^2 \right) }{\alpha}}.
    \end{eqnarray}
    Then,
    \begin{eqnarray}
        \left| \mathbf{E} [ \Lambda' ] - \Lambda \right| & \leq & \sqrt{ \frac{\log \left( \alpha  \left| S \right|^2 \right) }{\alpha}} + 2 \left| S \right| \exp \left( - \log \left( \alpha \left| S \right|^2 \right) /2 \right) \\
        & =  & \sqrt{ \frac{\log \left( \alpha  \left| S \right|^2 \right) }{\alpha}} +
        \frac{2}{\sqrt{\alpha}} \\
        & = & \frac{2 + \sqrt{ \log \alpha + 2 \log \left| S \right|}}{\sqrt{\alpha}},
    \end{eqnarray}
    as desired.
\end{proof}

\bibliographystyle{plain}

\bibliography{main}

\end{document}